\RequirePackage[l2tabu, orthodox]{nag}
\documentclass{stsci_report}
\usepackage[pdftex,
pdfauthor={D. Stark},
pdftitle={Using repeat imaging of a galaxy cluster taken over a seventeen-year baseline, we examine the impact that degraded Charge Transfer Efficiency (CTE) has on photometric measurements of extended sources using the ACS/WFC on HST. We examine how measured brightnesses depend on time since ACS installation, source location on the WFC detectors, source brightness, and local background level in individual exposures. We find that global brightness measurements using large apertures are generally reliable within $\sim$0.05 magnitudes across the WFC detectors if exposure backgrounds are above $20e^-/{pixel}$ and sources are brighter than $\sim300e^-$ in a single exposure. However, brightness measurements on smaller scales can suffer deficiencies in excess of 0.1 mags (sometimes, significantly more) in recent data unless sources are very close to the CCD serial registers ($<512$ pixels), or brighter than $\sim3000\,e^-$ in a single exposure. We also show how degraded CTE can result in artificial asymmetries in galaxy light distributions, which are largely mitigated if backgrounds are $>20e^-/{pixel}$ and targets are not far ($>1536$ pixels) from the serial registers. As expected, brightness measurements in later epoch data are best when using CTE-corrected images (FLC/DRC), but our results imply that the pixel-based CTE correction algorithm employed by the ACS reduction pipeline does not necessarily place charge back into its original location within extended sources. Based on this study, users are advised to keep backgrounds above the already recommended $30e^-/\mathrm{pixel}$, ensure targets will have at least $\sim 300e^-$ in a single exposure, and place targets close to the serial registers if analysis of their spatially resolved properties is needed.},
pdfkeywords={ACS, CCD, keywords}]{hyperref}
\copyrighttext{Copyright \copyright 2025 The Association of Universities for Research in Astronomy, Inc. All Rights Reserved.}
\usepackage{booktabs}
\usepackage{rotating}
\usepackage{natbib}
\usepackage{url}
\usepackage{graphicx}
\usepackage[font=footnotesize,labelfont=bf]{caption}
\usepackage{subcaption}
\usepackage{nameref}
\usepackage{hyperref} 
\usepackage{array} 
\usepackage{color}
\usepackage{amsmath, amssymb}
\usepackage{hhline}
\usepackage{multirow}
\usepackage{aas_macros}

\title{\textbf{The Impact of Degraded Charge Transfer Efficiency on Extended Sources in ACS/WFC}}
\presubtitle{\flushleft{\includegraphics[width=5cm]{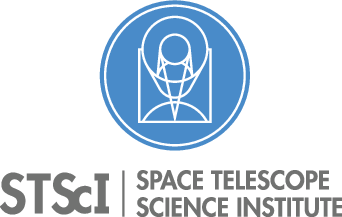}} \\ \hfill Instrument Science Report ACS 2025-02}
\author{D.~V. Stark, M. Chiaberge, N.~A. Grogin}
\date{October 7, 2025}

\begin{document}

\maketitle

\abstract{Using repeat imaging of a galaxy cluster taken over a seventeen-year baseline, we examine the impact that degraded Charge Transfer Efficiency (CTE) has on photometric measurements of extended sources using the ACS/WFC on HST. We examine how measured brightnesses depend on time since ACS installation, source location on the WFC detectors, source brightness, and local background level in individual exposures. We find that global brightness measurements using large apertures are generally reliable within $\sim$0.05 magnitudes across the WFC detectors if exposure backgrounds are above $20e^-/{pixel}$ and sources are brighter than $\sim300e^-$ in a single exposure. However, brightness measurements on smaller scales can suffer deficiencies in excess of 0.1 mags (sometimes, significantly more) in recent data unless sources are very close to the CCD serial registers ($\lesssim 512$ pixels), or brighter than $\sim3000\,e^-$ in a single exposure. We also show how degraded CTE can result in artificial asymmetries in galaxy light distributions, which are largely mitigated if backgrounds are $>20e^-/{pixel}$ and targets are not far ($>1536$ pixels) from the serial registers. As expected, brightness measurements in later epoch data are best when using CTE-corrected images (FLC/DRC), but our results imply that the pixel-based CTE correction algorithm employed by the ACS reduction pipeline does not necessarily place charge back into its original location within extended sources. Based on this study, users are advised to keep backgrounds above the already recommended $30e^-/\mathrm{pixel}$, ensure targets will have at least $\sim 300e^-$ in a single exposure, and place targets close to the serial registers if analysis of their spatially resolved properties is needed.}

\section*{Introduction} \label{s:intro}
An important property of Charge-Coupled Devices (CCDs) is their Charge Transfer Efficiency (CTE), which refers to how effectively charge is transferred between adjacent pixels during the read-out process. The Wide Field Channel (WFC) on the Advanced Camera for Surveys (ACS) aboard the Hubble Space Telescope (HST) has been in the harsh environment of space for over two decades, where high energy particles have damaged its CCD detectors, leading to degradation of its CTE, or an increase in its Charge Transfer Inefficiency (CTI). A decline in CTE leads to charge getting trapped and later released during the read-out process, leading to ``trails" behind sources in the direction opposite to the path taken by charge during read-out. This leads to targets having lower than expected brightness measurements and shifts in their centroids.

The degradation of CTE and its impact on the measured properties of point sources has been extensively studied over the lifetime of ACS. It has been shown that CTE is steadily declining with time, and has the largest negative impact when sources are faint, far from the serial registers (i.e., charge is transferred across many pixels), and on low ($<30 e^-/\mathrm{pixel}$) backgrounds \citep{Riess2004, Chiaberge2012, Chiaberge2022, Anderson2022b}. Based on these trends, empirical correction formulas for point-source brightness have been developed \citep{Chiaberge2012, Chiaberge2022}, and a pixel-based CTE correction algorithm, calibrated on individual hot pixels from the WFC detectors, was developed and is part of the standard WFC reduction pipeline to produce ``CTE-corrected" (FLC/DRC) images \citep{Anderson2010, Anderson2018}. 

However, the impact of degraded CTE on {\it extended} source photometry has not been studied until now. It is possible that extended sources suffer relatively less brightness loss from degraded CTE (compared to point sources) because their extended nature means they effectively raise the local ``background" around many of their component pixels, also known as ``self-shielding". Nonetheless, they still have numerous pixels adjacent to empty sky as well as faint regions which, depending on the background level and location on the detector, could still be subject to the effects of CTI. It is important to carefully test the impact of degraded CTE on extended sources to ensure reliable results can be derived from ACS/WFC imaging data for a variety of science cases. 

In this report, we present an analysis of the impact that degraded CTE has on the measured brightnesses of extended sources using imaging data from a galaxy cluster field that has been observed using the same filters over a seventeen year baseline. We directly compare brightnesses measured using matched large and small apertures, as well as the 2D light distributions, over a range of epochs. We test how brightness measurements depend on time since ACS installation, as well as the local backgrounds, brightnesses, and locations of objects in individual exposures. We conclude with a series of recommendations observers should follow to ensure optimal measurements with WFC. 

\section*{Data}
\begin{figure}
    \centering
    \includegraphics[width=\linewidth]{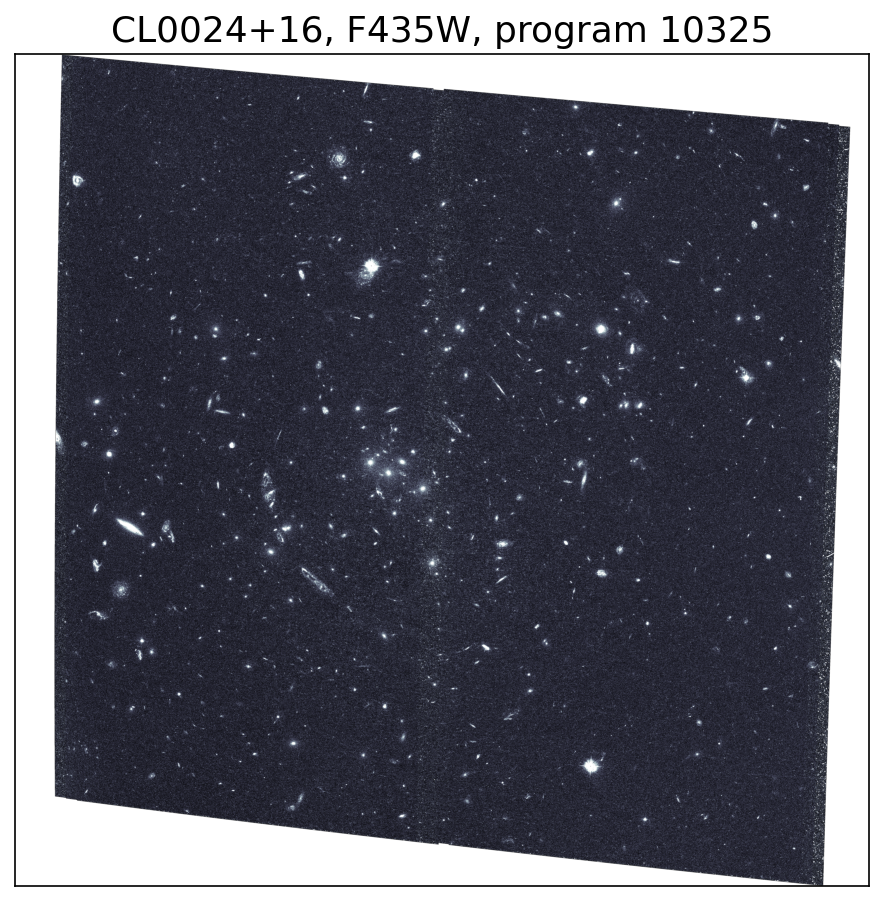}
    \caption{The galaxy cluster CL0024$+$16 used for this study, imaged in F435W as part of GO 10325.}
    \label{fig:field_full}
\end{figure}

WFC imaging data of galaxy cluster CL0024+16 is used for this study, and the observing programs used are summarized in Table~\ref{tbl:program_summary}. CL0024+16 was first observed in 2004 as part of HST GO 10325 (PI: H. Ford). A F435W image from this program is shown in Figure~\ref{fig:field_full}). Although these first observations were taken after the WFC CCDs had been subjected to the harsh environment of outer space for $\sim2$ years, the degradation of its CTE at that point is expected to be minimal. Additionally, these images are more than three times deeper than other imaging data for these fields, further reducing any impact of degraded CTE. Therefore,  we treat the 2004 data as ``truth" and use them as the reference for comparison to later epochs.

Observations of this same field were carried out in 2013 and 2021 (programs CAL/ACS 13603 and CAL/ACS 16870, PI: M. Chiaberge), after ACS had been in space for 11 and 19 years. These observations were explicitly designed to yield lower backgrounds in individual exposures in order to directly test situations where the effects of CTI are expected to be most pronounced. Therefore, a blue filter (F435W) was used, and exposure times were limited to $4\times 500$ seconds. ACS/CAL 13603 and ACS/CAL 16870 each consisted of three orbits, and used a tight four-point dither pattern in each orbit. The second and third orbits had LED post-flash durations of 0.7 and 1.4 seconds to increase the background levels by $\sim10$ and $\sim20e^-{\rm pix}$ over the intrinsic sky background of $\sim15-20e^-/{\rm pix}$ (averaged over the detector in the unflashed exposures)\footnote{The post-flash LED does not uniformly illuminate the CCDs, so background levels can vary by up to $\sim10e^-/\mathrm{pixel}$ across the detector in post-flashed exposures.}.

\begin{table}[]
    \centering
    \begin{tabular}{|c|c|c|c|c|c}
    \hline
    \textbf{program} & \textbf{Year} & \textbf{ASN} & \textbf{EXPTIME} & \textbf{FLASHDUR}\\
    \textbf{ID} &  & \textbf{ID} & \textbf{[s]} & [s]\\
    \hline
    GO 10325 & 2004 & J91JC4010 & $4 \times 1305 + 1\times 1215$ & 0 \\
    ACS/CAL 13603 & 2013 & JCH001010 & $4\times 500$ & 0 \\
    ACS/CAL 13603 & 2013 & JCH001020 & $4\times 500$ & 0.7 \\
    ACS/CAL 13603 & 2013 & JCH001030 & $4\times 500$ & 1.4 \\
    ACS/CAL 16870 & 2021 & JERU01010 & $4\times 500$ & 0 \\
    ACS/CAL 16870 & 2021 & JERU01020 & $4\times 500$ & 0.7 \\
    ACS/CAL 16870 & 2021 & JERU01030 & $4\times 500$ & 1.4 \\
    \hline
    \end{tabular}
    \caption{Summary of observing programs targeting CL0024+16 used in this study.}
    \label{tbl:program_summary}
\end{table}

\section*{Methods}

In the following section, we discuss how our combined images were generated, how these were aligned with one another, how sources were detected and their properties measured, and the final quality cuts applied to our data set.

\subsection*{Image Generation and Alignment}
Drizzlepac (specifically the \texttt{Astrodrizzle} Python routine) is used to generate combined distortion-free images for each program/orbit. We generate both non-CTE-corrected DRZ images and CTE-corrected DRC images. The following are the key parameters we set:
\begin{itemize}
    \item \texttt{pixfrac=0.7}
    \item \texttt{skymethod=globalmin+match}
    \item \texttt{final\_scale=0.03333}
    \item \texttt{final\_rot=0} 
    \item \texttt{driz\_cr\_snr = 3.0 2.5}
    \item \texttt{combine\_type = minmed} 
\end{itemize}
In particular, the values of \texttt{final\_scale} and \texttt{driz\_cr\_snr} were set smaller than default to enable more robust rejection of cosmic rays. Setting \texttt{combine\_type = minmed} was also very important for masking cosmic rays when only four exposures were available, otherwise several residual cosmic rays were present in the final image.  Defining \texttt{final\_rot=0} ensured all the images had the same angular orientation relative to north, but the centers of the fields were not identical. Since our goal was to conduct matched aperture photometry across all epochs and post-flash levels, these images needed to be aligned. There are only a few bright stars in the field with which to conduct alignment, but regardless, we did not want to use stars for alignment due to the potential that they moved appreciably in the span of years separating our imaging at different epochs. We therefore used phase cross-correlation on the full images, specifically using the \texttt{phase\_cross\_correlation} routine as part of the \texttt{skimage} Python package. In all cases, the 2004 DRC image was considered the reference image against which all other images were aligned to. 

To ensure any high proper motion stars did not skew our results, we used \texttt{SExtractor} \citep{Bertin1996} to identify likely stars in both the reference image and the image to be shifted, then replaced those regions with random noise consistent with the noise properties of the respective image, effectively masking out the stars. The images were upsampled by a factor of ten (implemented via an optional keyword in \texttt{phase\_cross\_correlation}) to improve the accuracy of the alignment. Ideally, we would mask additional regions we did not want considered during the alignment, e.g., regions with no coverage like the chip gap. Unfortunately, the \texttt{phase\_cross\_correlation} routine disables upsampling if such masks are used, lowering the accuracy of the alignment. Therefore, we conducted the alignment using a large contiguous region, slightly less than 50\% of the field, that avoided the chip gap in all images. Lastly, to avoid CTE trails artificially skewing the alignment, we use the DRC image from each program/orbit for the alignment and apply the same calculated shift to the corresponding DRZs. 

\begin{figure}
    \centering
    \includegraphics[width=\linewidth]{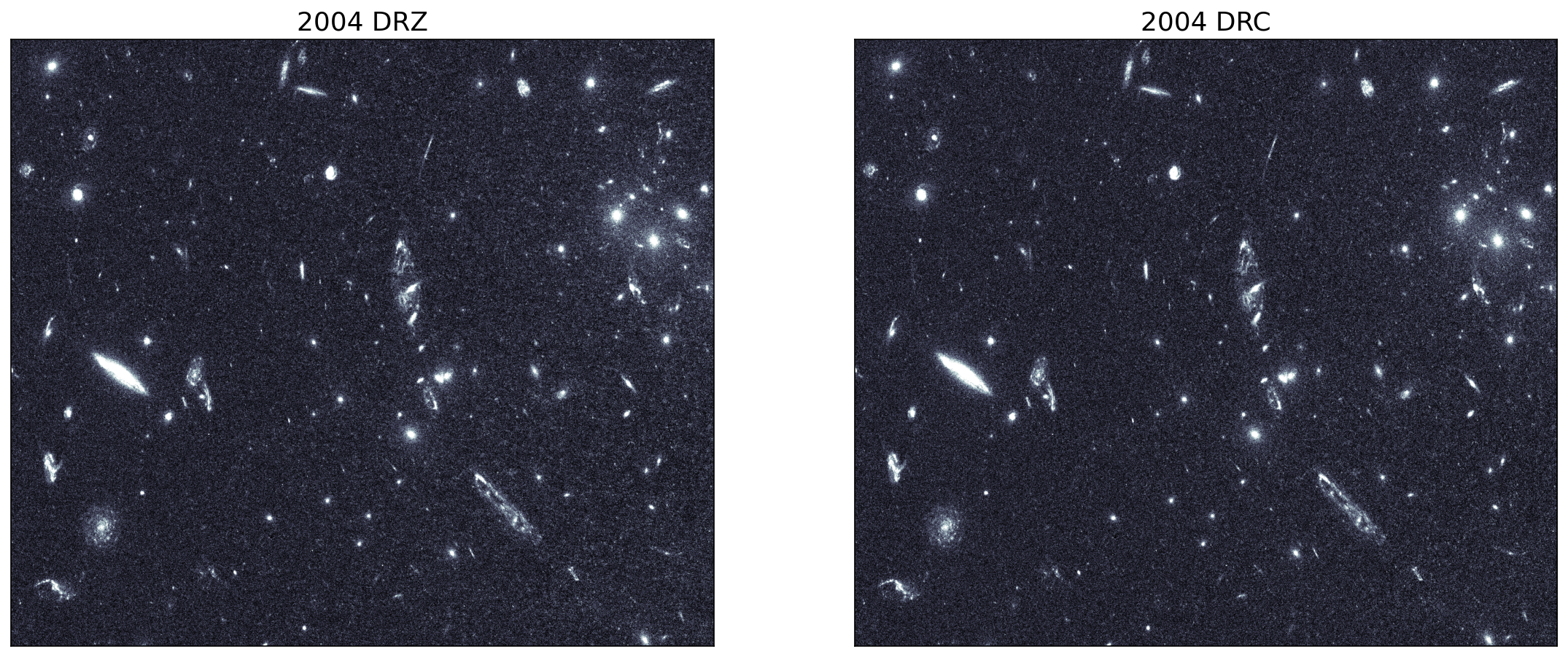}
    \includegraphics[width=\linewidth]{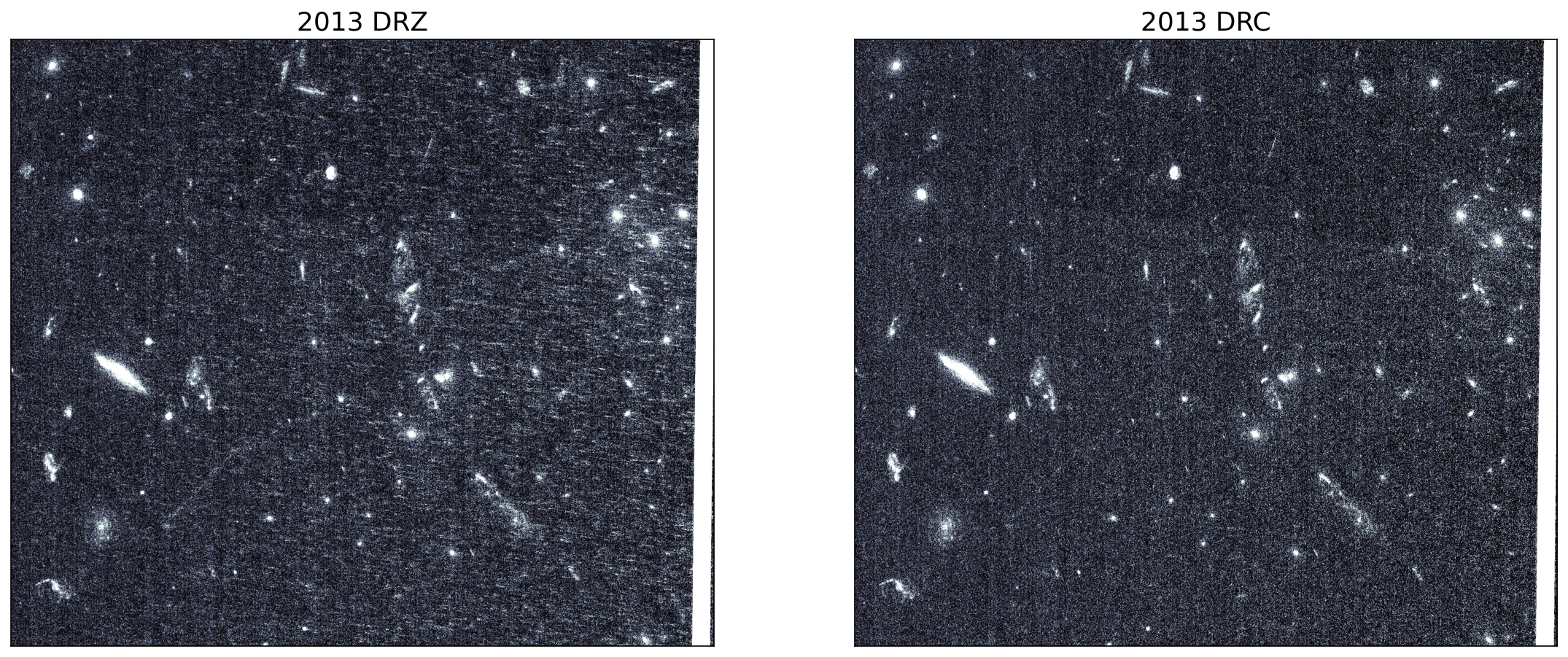}
    \includegraphics[width=\linewidth]{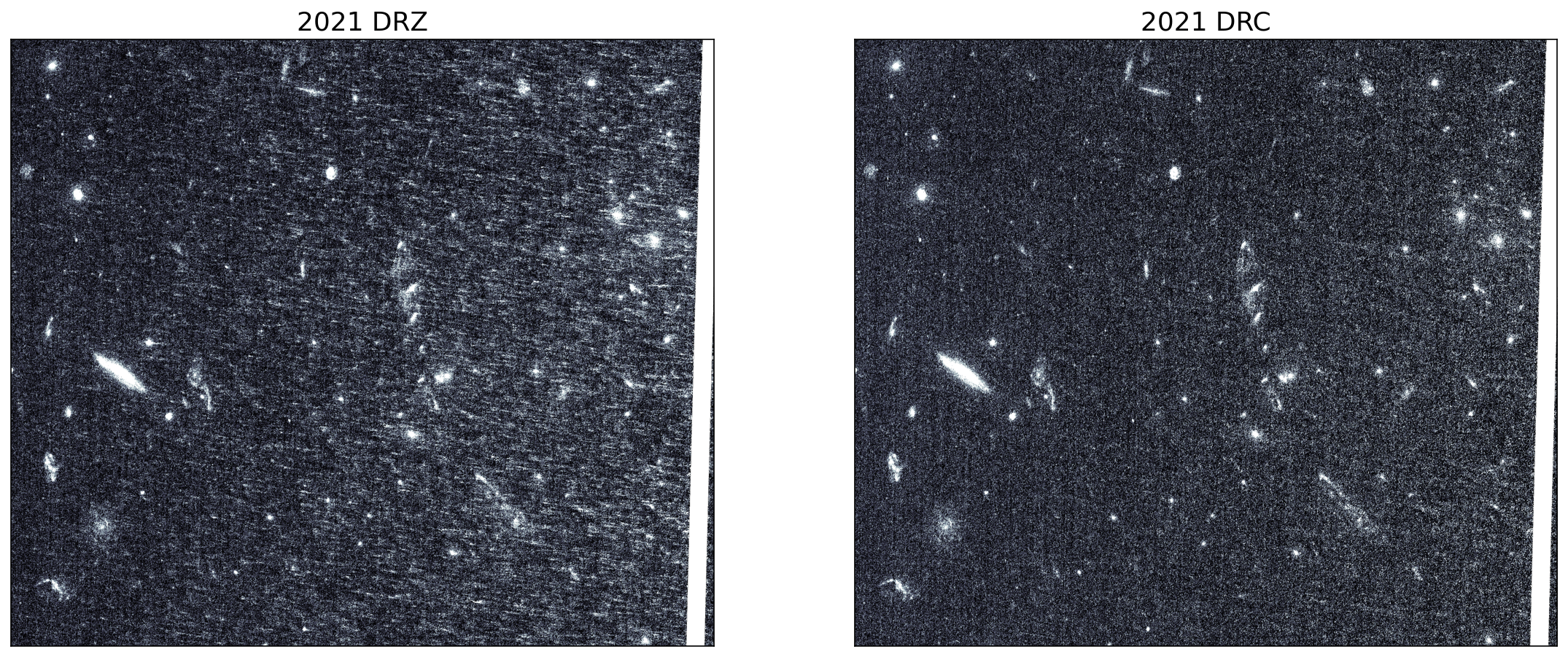}
    \caption{The same zoomed in region of sky seen in images from the three epochs used in our analysis. Both DRZ (left) and DRC (right) images are shown. No post-flashed images are shown. The scaling is identical between all images. The white vertical band seen in the 2013 and 2021 images is the WFC chip gap which was not covered by the 4-point dither pattern.}
    \label{fig:zoomfield_3epochs}
\end{figure}

Figure~\ref{fig:zoomfield_3epochs} shows the same zoomed in region from the aligned DRZ and ``CTE-corrected" DRC images across all three programs used for this study. Only the images without post-flash are shown. The impact of degraded CTE is apparent in the 2013 and 2021 DRZ images as seen by the CTE ``trails" that grow more pronounced as the distance from the serial register increases (approximately corresponding to the left-to-right direction in these cutouts). The DRC images show significantly less CTE trail structure.

\subsection*{Source Detection and Characterization}
We used the \texttt{photutils} Python package as the primary tool to detect and measure properties of sources in this study. The overall strategy was to treat the deeper and minimally CTE-impacted 2004 DRC image as the ``truth" image and use it to detect sources and measure their shape parameters. These shape parameters were used to define elliptical apertures for brightness measurements. We applied the {\it exact same} apertures to the already aligned later-epoch data in order to extract aperture-matched photometry across all imaging data for as fair of a comparison as possible. In the remainder of this section, we describe in detail the steps taken to extract these measurements.

First, any large-scale residual background was subtracted from each image\footnote{\texttt{photutils.background.Background2D}}. A box size of $512\times512$ pixels was used. This relatively large box size was chosen to ensure individual galaxies did not significantly influence the background estimation, although we found the specific choice of box size did not significantly impact the number or properties of detected sources. The background level was determined using the same methodology as SExtractor to estimate the mode of the image in each box. A $3\times3$ median filter was then applied to the low-resolution background map, after which it was spline-interpolated onto the original image grid. Simultaneously, a map of the background rms noise level was calculated by sigma-clipping values more than 3$\sigma$ from the median using up to ten iterations.

Source detection was then performed on the background-subtracted 2004 image\footnote{\texttt{photutils.segmentation.detect\_sources}}. The data was first convolved with a Gaussian kernel with full width at half maximum (FWHM) of 4 pixels. We required sources to have a minimum of twenty-five connected pixels above a threshold of 2 times the rms noise level, where the local noise level was taken from the spatially resolved rms noise map described above. We did not attempt to deblend the sources identified via this procedure, as we found it tended to shred the galaxies in our field, especially because we were using a short wavelength filter where galaxy light tends to be clumpier. This choice means there are some blended galaxies in our data, but only a small fraction of objects was impacted, and blended galaxies are still extended sources (although perhaps with unusual shapes).

The segmentation map generated in the previous step was used to create a detection catalog\footnote{\texttt{photutils.segmentation.SourceCatalog}} that included source positions, shape properties, and aperture photometry. Galaxy positions and shape properties (radius, position angle, ellipticity) were estimated by analyzing the moments of each detected source, similar to \citet{Bertin1996}. We measured integrated brightnesses of our sources using elliptical apertures with radii of $kR_{\rm Kron}$, where $R_{\rm Kron}$ is the Kron radius (i.e., a light-weighted mean radius of all pixels in the image; \citealt{Kron1980}), and k is a scale factor. We used a scale factor of $k=2$ to capture the ``total" brightness of our targets. Technically, \citealt{Kron1980} argues this scale factor should include $\sim 90\%$ of galaxy light, but we chose not to increase the scale factor further to avoid adding significant noise to our measurements from extremely faint pixels at large radii. We also estimated central brightnesses using a scale factor of $k=0.5$. In cases where these elliptical apertures had masked data due to a nearby source or missing coverage, masked pixels were replaced with values from the opposite side of the source. Local background subtraction was also performed around each target using a rectangular annulus 32 pixels wide. For each source, the inner edge of this annulus was defined to be 1.5 times larger than maximum extent of the corresponding segmentation map. As a final step, \texttt{SExtractor} was run on the 2004 image to help differentiate between stars and galaxies using the \texttt{CLASS\_STAR} output. The \texttt{SExtractor} catalog was cross-matched with the \texttt{photutils} catalog using a match radius of one pixel. 

\begin{figure}
    \centering
    \includegraphics[width=0.75\linewidth]{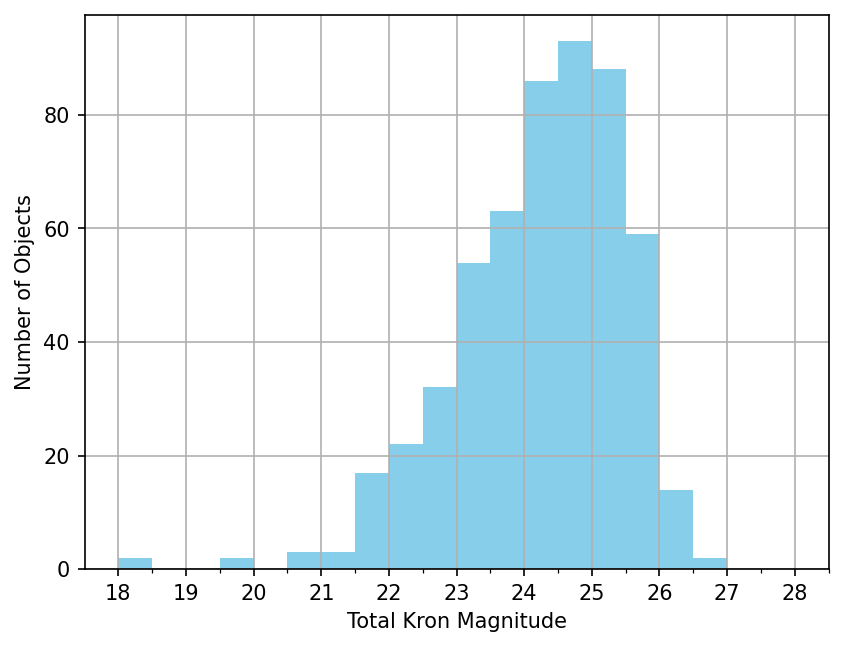}
    \includegraphics[width=0.75\linewidth]{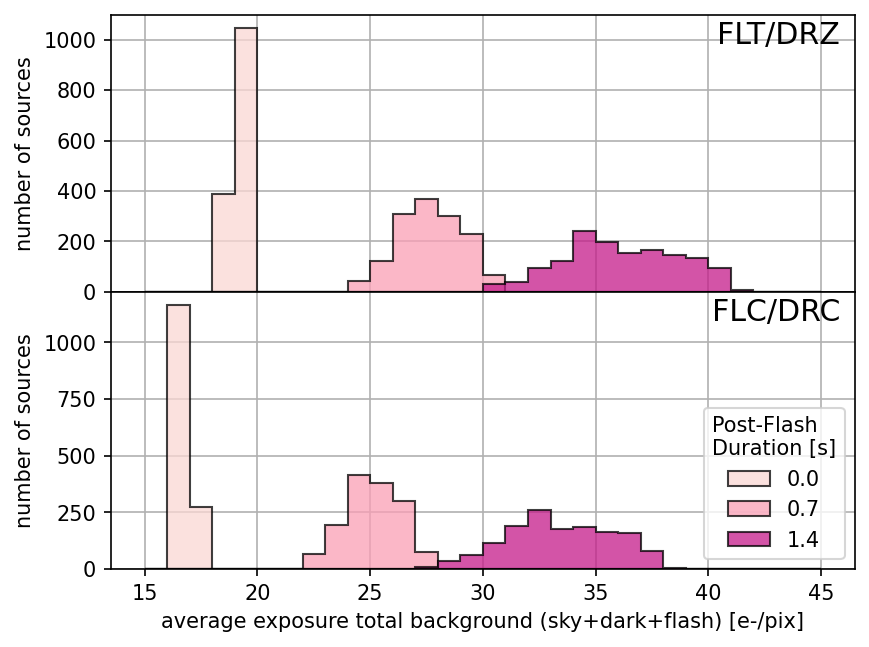}
    \caption{(top) Total Kron Magnitude distribution of detected sources from the 2004 data set with total Kron $SNR>10$. (bottom) Total background level (including sky, dark current, and post-flash) measured in individual exposures of the 2021 imaging data at the positions of detected sources, broken up by the image post-flash duration. Due to the non-uniformity of the flash, post-flashed data span a range of backgrounds. The backgrounds in the 2013 data span a comparable range.}
    \label{fig:mag_background_hist}
\end{figure}

Magnitudes were calculated in the STMAG system using
\begin{equation}
    m = -2.5\log{\left(f \times PHOTFLAM\right)}-21.1
\end{equation}
where $f$ is the measured brightness in counts per second and $PHOTFLAM$ is the inverse sensitivity in erg cm$^{-2}$ $\AA^{-1}$ electron$^{-1}$ reported in each image header. For simplicity, the magnitudes calculated using Kron scale factors of 2 and 0.5 are referred to as ``total magnitudes" and ``central magnitudes" throughout this report. Figure~\ref{fig:mag_background_hist} shows the distribution of total Kron magnitudes measured from the 2004 data for all sources detected with signal-to-noise ratio (SNR; simply defined as the ratio of total brightness divided by its uncertainty) greater than 10. 

Brightness measurements for the later epoch mages were taken by applying the exact same apertures defined using the 2004 image. Again, a local background subtraction was performed around each source using identical rectangular annuli. We also extracted information about the background at each source's location {\it in the original exposures}, which is a combination of the sky background, dark current, and post-flash (if any). To obtain sky backgrounds in each FLT/FLC exposure we used 2D background maps calculated in a similar manner as for our final combined images (see above) but using a box size of $256\times 256$ pixels. The dark and post-flash background maps were obtained by scaling the reference flash and dark files by the respective exposure time of 500 seconds. Since dither patterns were used for these observations, each source technically lied at a slightly different location on the detector during each exposure. However, the dither patterns were tight and these variations small, so for each source, we measured the local background in each of the four exposures and then took the average.  Figure~\ref{fig:mag_background_hist} shows the distribution of these background values measured in the 2021 images. We sample exposure backgrounds ranging from $\sim15$ to $\sim 40e^-/\mathrm{pixel}$. The images with flash durations $>0s$ show a spread of backgrounds primarily due to the non-uniformity of the post-flash LED. 

\subsection*{Data Quality Selection}
Unless stated otherwise, our analysis uses the following selections on the data:
\begin{itemize}
    \item Total Kron SNR$>$ 10 in the 2004 detection image
    \item Total Kron SNR$>3$ in the 2013 and 2021 images
    \item Total Kron magnitude between 21.5 and 26. Beyond these bounds the data are poorly sampled.
    \item Stellarity index (from the \texttt{SExtractor} \texttt{CLASS\_STAR} parameter) $<0.3$\footnote{The overwhelming majority of objects had \texttt{CLASS\_STAR}$< 0.02$ or $>0.95$. Our results are not sensitive to the specific cut in \texttt{CLASS\_STAR} as long as the locus of points with high values are rejected.}
    \item Objects around the outer edges and very center of the 2004 image (where the chip gap is covered but at less depth) are rejected. The shallower depth in these regions leads to an increase in spurious detections. While most are rejected via SNR cuts, we reject the entire regions to ensure a clean sample.
\end{itemize}

\section*{Results}
In the following section we show how brightness measurements have changed as a function of time in our imaging fields. Most of the results will display the difference in magnitude between two epochs defined as:
\begin{equation}
    \Delta m = m_{measured} - m_{truth}
\end{equation}
where $m_{truth}$ is the ``true" magnitude measured from the 2004 data, and $m_{measured}$ refers to the magnitude measured in the later epoch data. A value of $\Delta m = 0$ indicates consistent brightness measurements in both epochs, while values $\Delta m > 0$ or $\Delta m < 0$ indicate deficient or excess brightness in later epochs. We analyze $\Delta m$ with respect to three other variables: the average background level per exposure, the brightness per exposure, and the average number of $y$ pixel transfers for a source to be read-out. If degraded CTE impacts extended source brightness measurements, it is expected that $\Delta m$ will increase with increasing $y$ pixel transfers, decreasing exposure background, and decreasing source brightness.

\subsection*{Global photometry} \label{sec:global_phot}

\begin{figure}
    \centering
    \includegraphics[width=\linewidth]{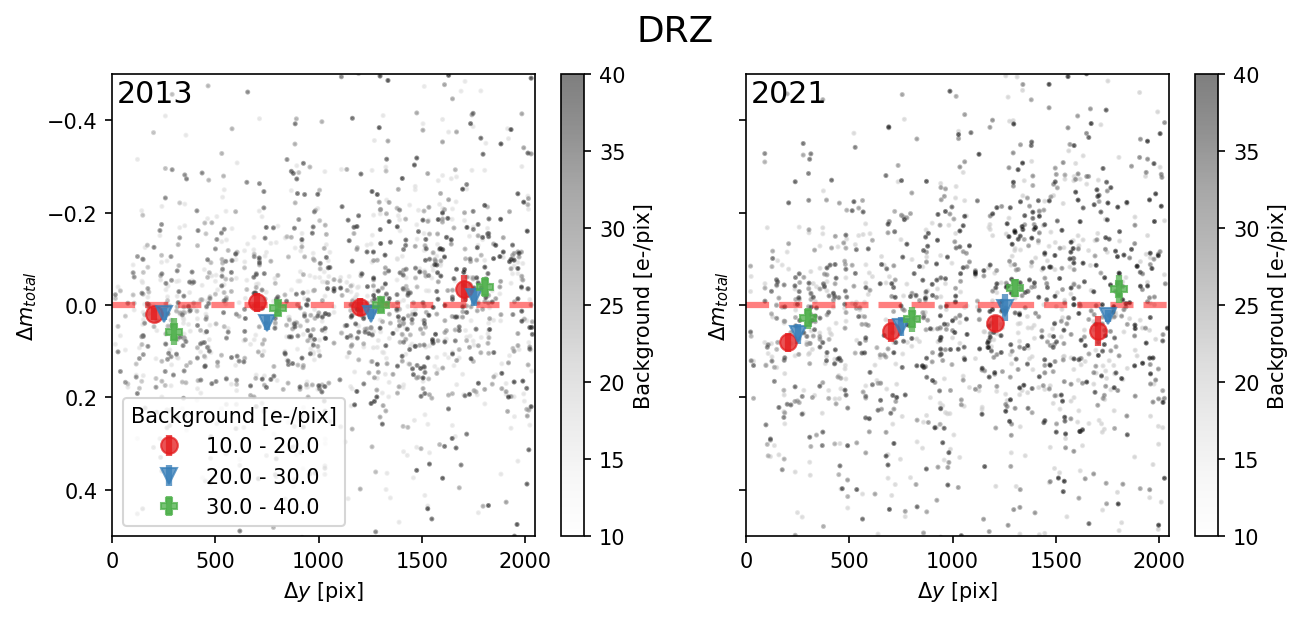}
    \includegraphics[width=\linewidth]{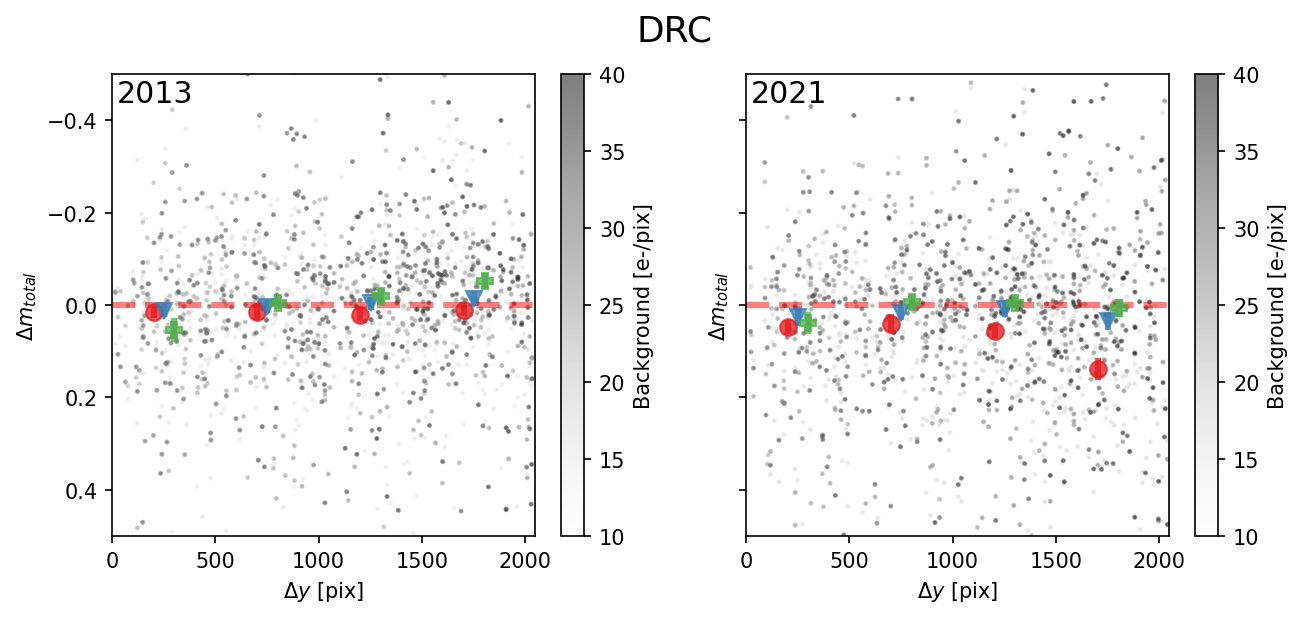}
    \caption{$\Delta m_{total}$ versus $\Delta y$ for DRZ (top) and DRC (bottom) images taken in 2013 (left) and 2021 (right) using total Kron magnitudes. Points are color-coded by the local background of each source the original exposures. Large points show binned median values as a function of $\Delta y$ and background. Global brightness measurements are generally reliable to within $\lesssim 0.05$ mags except when backgrounds fall below $20e^-/\mathrm{pixel}$ in more recent data.}
    \label{fig:dmag_yshifts_background_kron2}
\end{figure}

Figure~\ref{fig:dmag_yshifts_background_kron2} shows $\Delta m$ versus the mean number of $y$ pixel transfers needed to read-out a target (given the shorthand $\Delta y$) for total Kron magnitudes measured in the 2013 (left) and 2021 data (right) and for both DRZ (top) and DRC (bottom) images. We give this version of $\Delta m$ the variable name of $\Delta m_{total}$. Individual small data points represent individual galaxies. We have combined measurements across all post-flash levels, meaning individual unique galaxies will typically appear three times at three different background levels. Larger data points represent mean values as a function of $\Delta y$ and exposure background. Errors on these means are estimated from bootstrap resampling with 1000 iterations and replacement. Often these error bars are too small to be visible. Bootstrap resampling is chosen because it captures the uncertainty from both the random measurement errors, but also potential uncertainties due to random sampling of data in each bin. Generally, these uncertainties are consistent with simply propagating-forward the random uncertainties of the brightness measurements.

For the 2013 data, the mean value of $\Delta m_{total}$ across the whole detector is less than 0.03 at all backgrounds, and for both DRZ and DRC measurements. Often the values of $\Delta m_{total}$ are consistent with zero within the error bars. There are some weak correlations with $\Delta y$ such that $\Delta m_{total}$ is actually decreasing with $\Delta y$ at higher background levels, but these correlations are marginal at between $2-3\sigma$ significance. In the 2021 data, values of $\Delta m_{total}$ remain $<0.04$ for the DRZ and DRC measurements as long as backgrounds are $>20e^-/{\rm pix}$. Below backgrounds of $20e^-/\mathrm{pixel}$, mean $\Delta m_{total}$ increases to $\sim$0.06 and $\sim$0.1 for the DRZ and DRC measurements, respectively, and the DRC $\Delta m_{total}$ measurements show a clear correlation with $\Delta y$. Again, at the highest backgrounds we detect $2-3\sigma$ significance anticorrelations between $\Delta m_{total}$ and $\Delta y$ in the DRZ measurements, but these disappear when using DRC measurements. The immediate cause of these correlations, if real, is unclear. They may be related to CTE trail structure, residual post-flash background, or both.

These results show that, with the exception of data at backgrounds $<20e^-/\mathrm{pixel}$, the degradation of CTE in the WFC does not have a major impact on global extended source brightness measurements. Any average systematic offset between data at different epochs should generally be $\lesssim0.05$ mags.

\subsection*{Central photometry} \label{sec:central_phot}
\begin{figure}
    \centering
    \includegraphics[width=\linewidth]{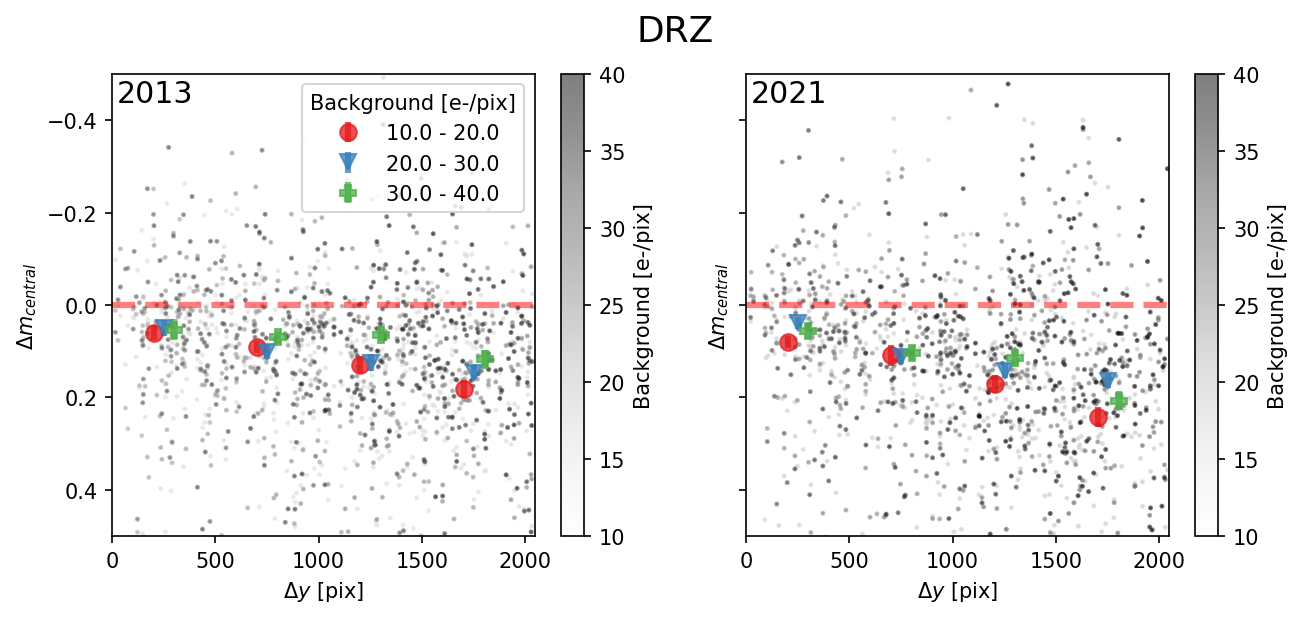}
    \includegraphics[width=\linewidth]{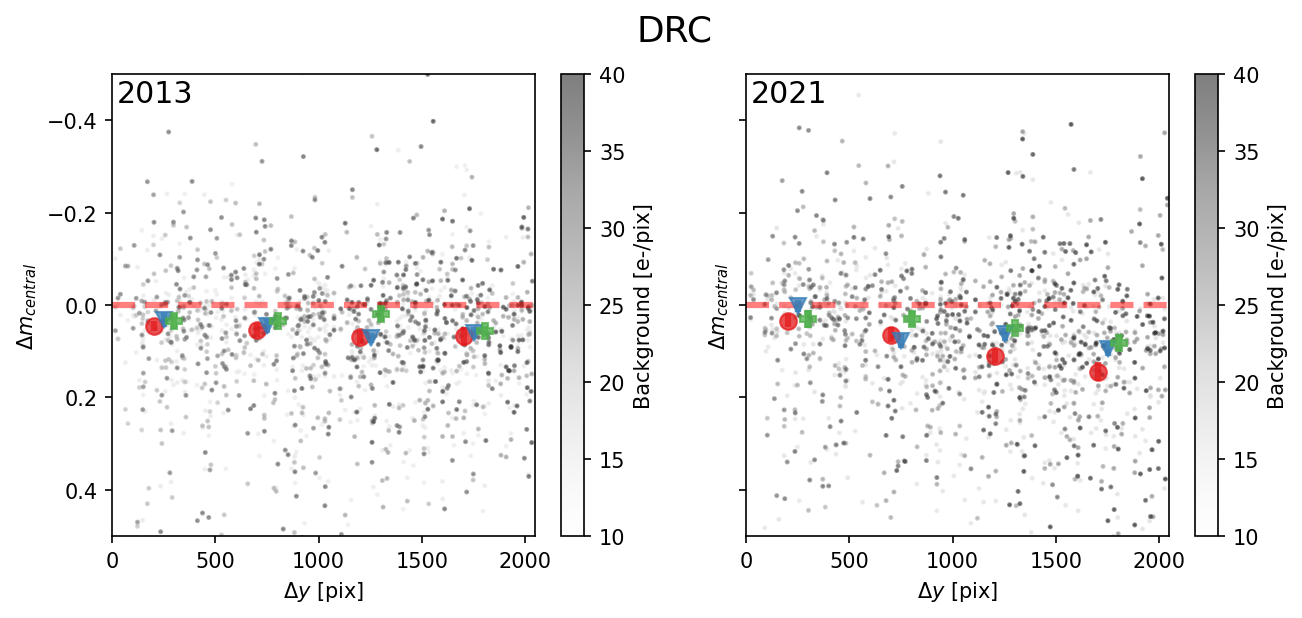}
    \caption{$\Delta m_{central}$ versus $\Delta y$ for DRZ (top) and DRC (bottom) images using apertures covering only the centers of galaxies. Unlike for global brightness measurements, there is a clear decline in recovered central brightness with increasing time, increasing $\Delta y$, and decreasing background.}
    \label{fig:dmag_yshifts_background_kron0.5}
\end{figure}

Figure~\ref{fig:dmag_yshifts_background_kron0.5} shows the same analysis described the previous section except now $\Delta m$ is calculated using central magnitudes only ($\Delta m_{central}$). Compared to total brightness measurements, the central brightness measurements show more significant mean offsets and often strong correlations with $\Delta y$ indicative of degraded CTE.

For the 2013 imaging data, the average $\Delta m_{central}$ over the entire FOV ranges from 0.09-0.14 and 0.05-0.09 magnitudes for the DRZ and DRC measurements, respectively. In the 2021 data, the mean offsets increase to 0.014-0.17 or 0.07-0.11 magnitudes for the DRZ and DRC measurements. In all cases, the offsets are larger when backgrounds are lower. Notably, although the pixel-based CTE correction used to generate DRC images lowers the mean value of $\Delta m_{central}$ across the detector, it does not remove it entirely. 

In the 2013 DRZ data, there are strong ($>3\sigma$) correlations between $\Delta m_{central}$ and $\Delta y$ for backgrounds $<30e^-/{\rm pix}$ and a marginal ($>2\sigma$) correlation at higher backgrounds. The value of $\Delta m_{central}$ increases with $\Delta y$ consistent with expectations of degraded CTE. The slope $\Delta m_{central}/\Delta y$ is roughly twice as large for the sources at backgrounds of $10-20e^-/{\rm pix}$ compared to the sources at $30-40e^-/{\rm pix}$.  In the DRC images, these correlations are weakened and made completely negligible for backgrounds $>20e^-/{\rm pix}$, although as noted above, a non-zero mean $\Delta m_{central}$ remains. 

In the 2021 data the correlations with $\Delta y$ are stronger. In the DRZs, correlations between $\Delta m_{central}$ and $\Delta y$ are all statistically significant with slopes that are 1.5-2 times steeper than in the 2013 data. Unlike in the 2013 data, correlations with $\Delta y$ remain even in the DRC data, although the slopes are shallower by roughly 50\%.

This analysis shows that extra caution must be taken when interpreting brightness measurements on local scales within galaxies across different epochs. In the very worst cases (low backgrounds, no CTE correction), differences can exceed 0.2 magnitudes. However, for sources close to the serial registers ($\Delta y<500-1000$ pixels, depending on the background), systematic offsets are $<0.05$ magnitudes.

\subsection*{Dependence on source brightness}
In our earlier analysis, we did not consider how $\Delta m$ may depend on source brightness, but it is expected that weaker sources will suffer from degraded CTE more than brighter sources. The relevant brightness, when it comes to CTI-related effects, is the number of counts in a given object per exposure, i.e., the fluence. We calculate this value, $f_{500}$, for every object by multiplying the counts per second in the combined drizzled image by the 500 second exposure time.

\begin{figure}
    \centering
    \includegraphics[width=\linewidth]{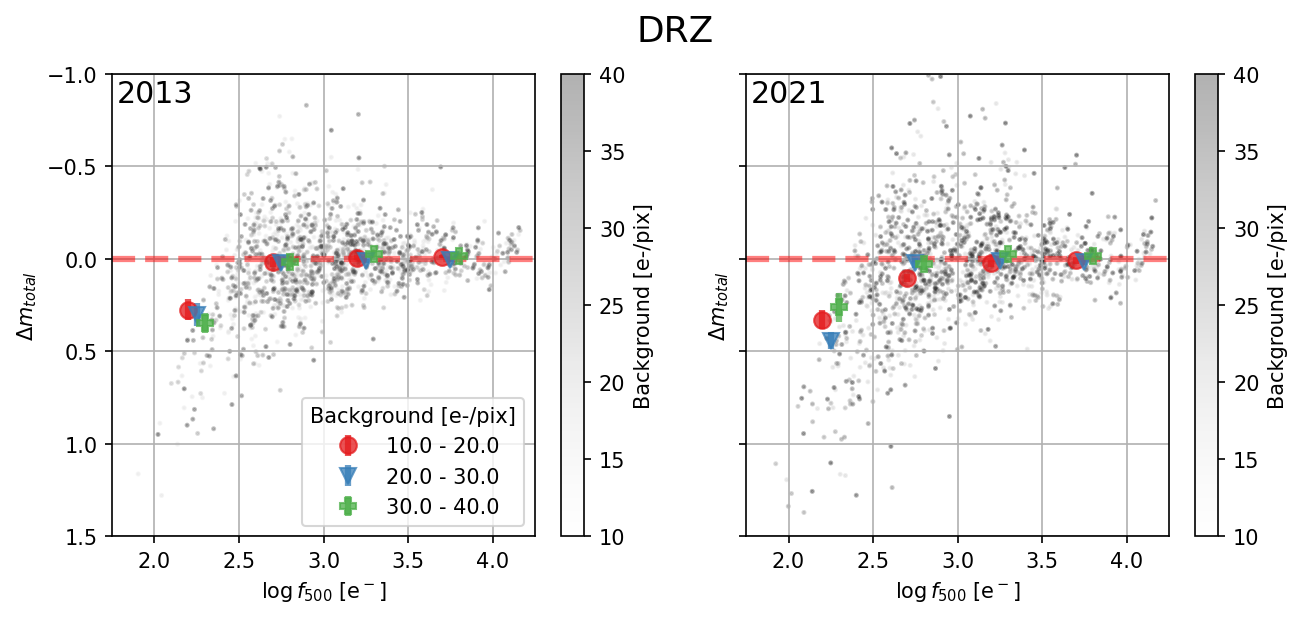}
    \includegraphics[width=\linewidth]{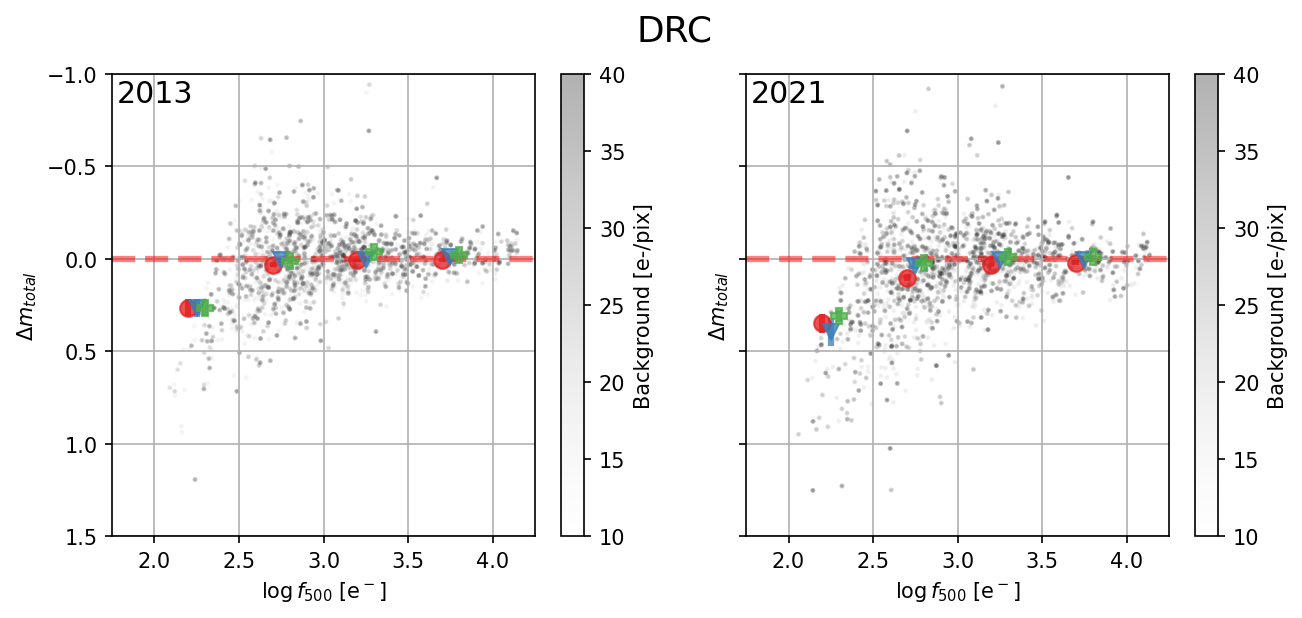}
    \caption{$\Delta m_{total}$ vs $f_{500}$, the counts per 500s exposure (fluence). Results from FLT/DRZ and FLC/DRC imaging data are shown on the top and bottom respectively. Brightness measurements become significantly underestimated for objects with total $\log{f_{500}} \lesssim 2.5e^-$, or $f_{500}\lesssim300e^-$, regardless of background level.}
    \label{fig:dmag_fluence}
\end{figure}

Figure~\ref{fig:dmag_fluence} shows $\Delta m_{total}$ vs $f_{500}$ for FLT/DRZ (top) and FLC/DRC (bottom) imaging data. For now, we do not separate sources as a function of $\Delta y$. In both epochs there is a clear relationship between $\Delta m$ and $f_{500}$ such that weaker sources show larger brightness deficits. Losses are mildly more pronounced at lower backgrounds and later epochs, but regardless of epoch and background, the rate at which $\Delta m_{total}$ increases grows significantly below $\log{f_{500}}\sim2.5$ or $\sim 300\,e^-$. This behavior is seen in the CTE-corrected data as well, implying that the losses are too significant, or the sources too weak, for the pixel-based correction to be effective. In fact, the pixel-based correction is purposefully designed to {\it not} try to correct brightnesses close to the read-noise level of an image in order to avoid noise amplification.

\begin{figure}
    \centering
    \includegraphics[width=\linewidth]{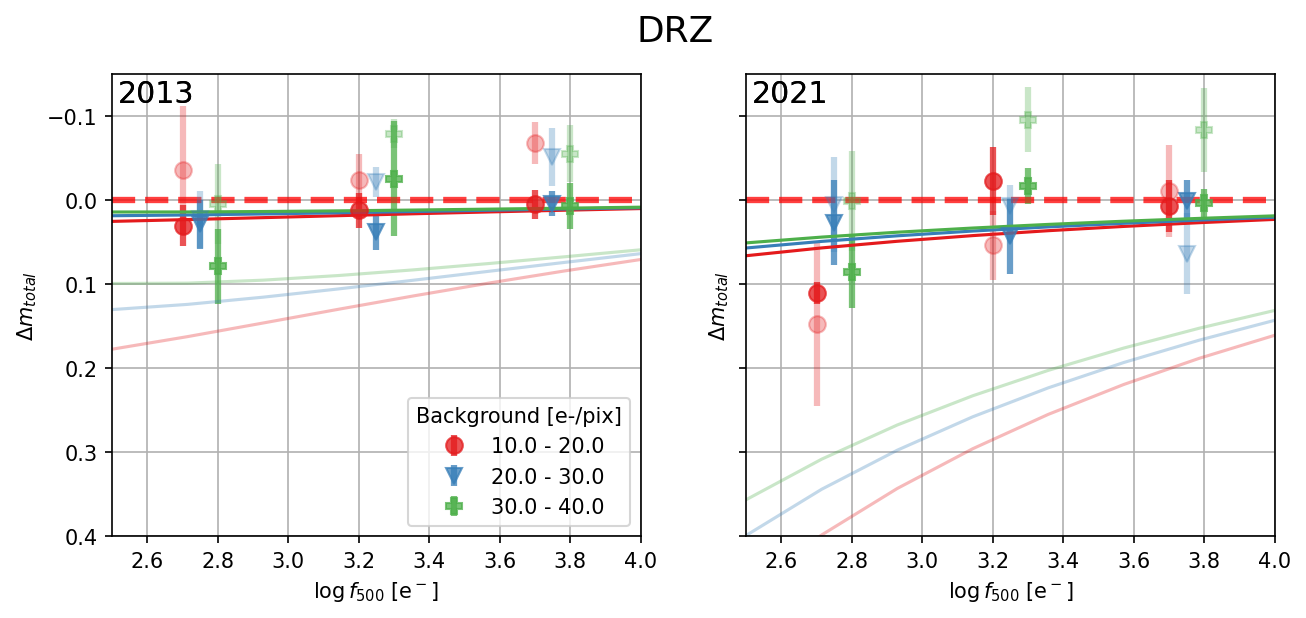}
    \includegraphics[width=\linewidth]{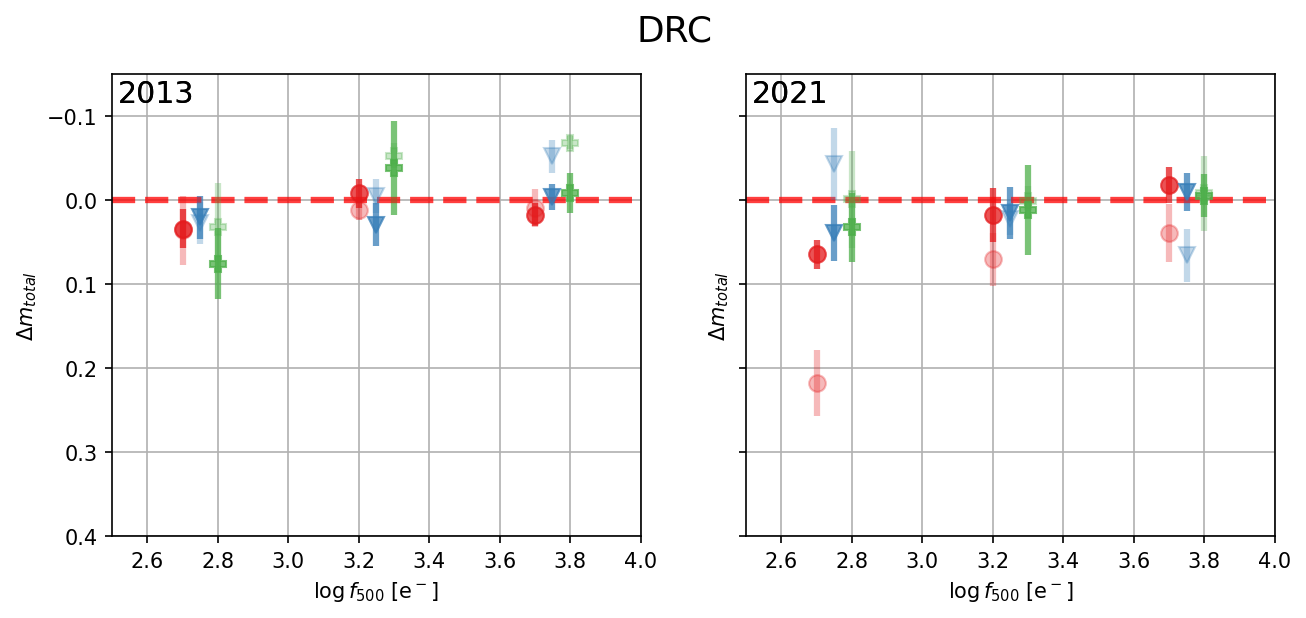}
    \caption{Average $\Delta m_{total}$ as a function of $f_{500}$, background, and $\Delta y$. Darker shaded points correspond to $\Delta y < 512$ pixels and lighter shaded points correspond to $\Delta y > 1536$ pixels. Darker and lighter solid lines correspond to the photometric CTE correction model of \citep{Chiaberge2022} for {\it point sources} at $\Delta y < 512$ and $\Delta y > 1536$, respectively, which are shown for comparison. Total brightness measurements are reliable to within $\sim0.1$ mag except in more recent data for objects far from the serial registers with $\log{f_{500}}<3$ and background $<20e^-/\mathrm{pixel}$.}
    \label{fig:dmag_dy_fluence_kronfactor2}
\end{figure}

 In Figure~\ref{fig:dmag_dy_fluence_kronfactor2} we plot the average value of $\Delta m_{total}$ as a function of all three relevant variables: $f_{500}$, background, and $\Delta y$. For clarity, we do not plot measurements for individual galaxies, and we only plot mean values at $\Delta y < 512$ and $\Delta y > 1536$ corresponding to the closest and furthest quadrants from the CCD serial registers. We have not plotted the mean values for objects with $\log{f_{500}}<2.5\,e^-$ or $\log{f_{500}}>4\,e^-$ due to the small number of data points per bin at those brightness levels when dividing the data by so many variables.

The 2013 DRZ and DRC measurements at low $\Delta y$ have $\Delta m_{total}$ averaged over all brightnesses less than $0.03$ mags across all background levels. There are no statistically significant trends with $f_{500}$. At higher $\Delta y$,  brightness tends to be slightly overestimated ($\Delta m < 0$), especially at $f_{500}>3\,e^-$, although $\Delta m$ never falls below -0.1 at any point. However, there are technically still no statistically significant correlations between $\Delta m$ and $f_{500}$ at any background level. 

The 2021 DRZ measurements show similar results as the 2013 measurements as long as $\log{f_{500}}>3$ and backgrounds $>20e^-/{\rm pix}$. At $\log{f_{500}} < 3\,e^-$ and backgrounds $<20e^-/{\rm pix}$, $\Delta m$ exceeds 0.1 magnitudes. In the DRC images, at low $\Delta y$, these measurements improve, but at high $\Delta y$, $\Delta m$ remains large ($>0.2$ magnitudes). Independent of $\Delta y$, the measurements with background $<20e^-/{\rm pix}$ show statistically significant correlations with $f_{500}$. For data at higher backgrounds, we do not find any statistically significant correlations between $\Delta m$ and $f_{500}$. 

For comparison, Figure~\ref{fig:dmag_dy_fluence_kronfactor2} also shows the corresponding values of $\Delta m$ for point sources using circular apertures with radii of 5 pixels, based on the model described in \citet{Chiaberge2022}. At low $\Delta y$, $\Delta m$ for point and extended sources are comparable and small. Notably, at large $\Delta y$, the brightness losses for point sources are significantly larger than for extended sources, by $\sim$0.1-0.3 magnitudes over the range of brightnesses and backgrounds shown. This supports the idea that extended sources are more ``self-shielded" from CTI-related losses, although as we have shown, they are not completely resistant to them.

The general conclusion from this analysis is that DRC measurements across the detector at backgrounds $>20e^-/{\rm pix}$ are reliable within a systematic error of $\sim0.05$ mags down to $\log{f_{500}}\sim2.5\,e^-$. At lower backgrounds, users should use caution when interpreting measurements of sources with $\log{f_{500}}<3\,e^-$, unless at low $\Delta y$. 

\begin{figure}
    \centering
    \includegraphics[width=\linewidth]{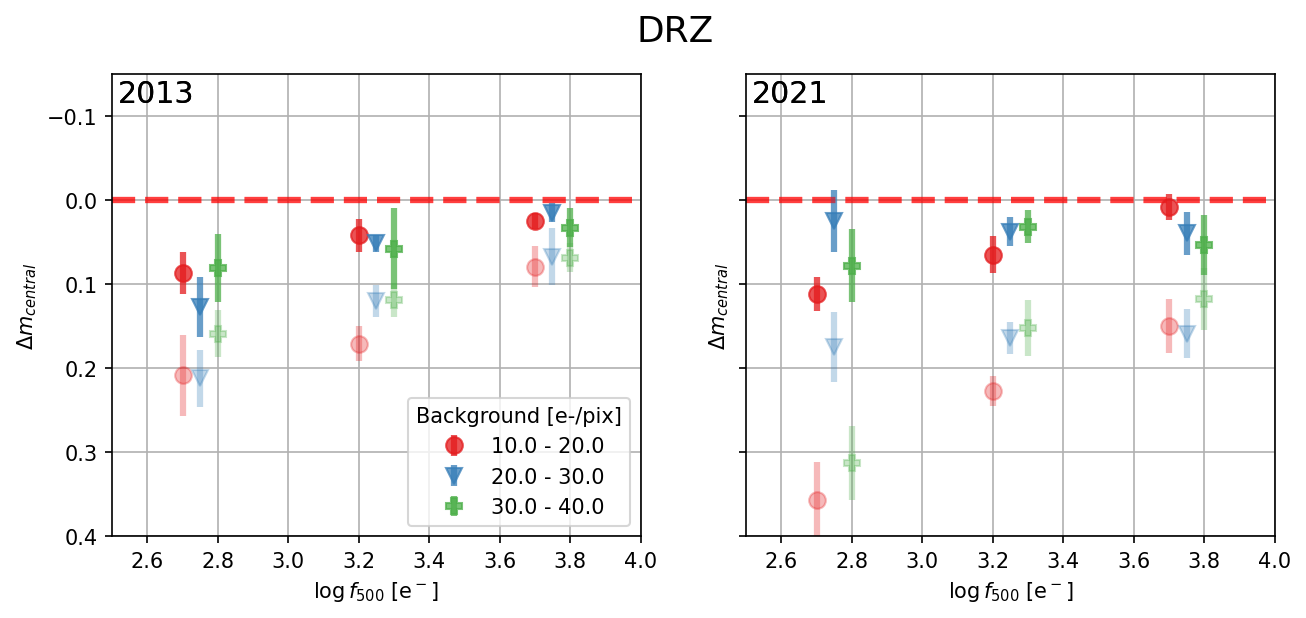}
    \includegraphics[width=\linewidth]{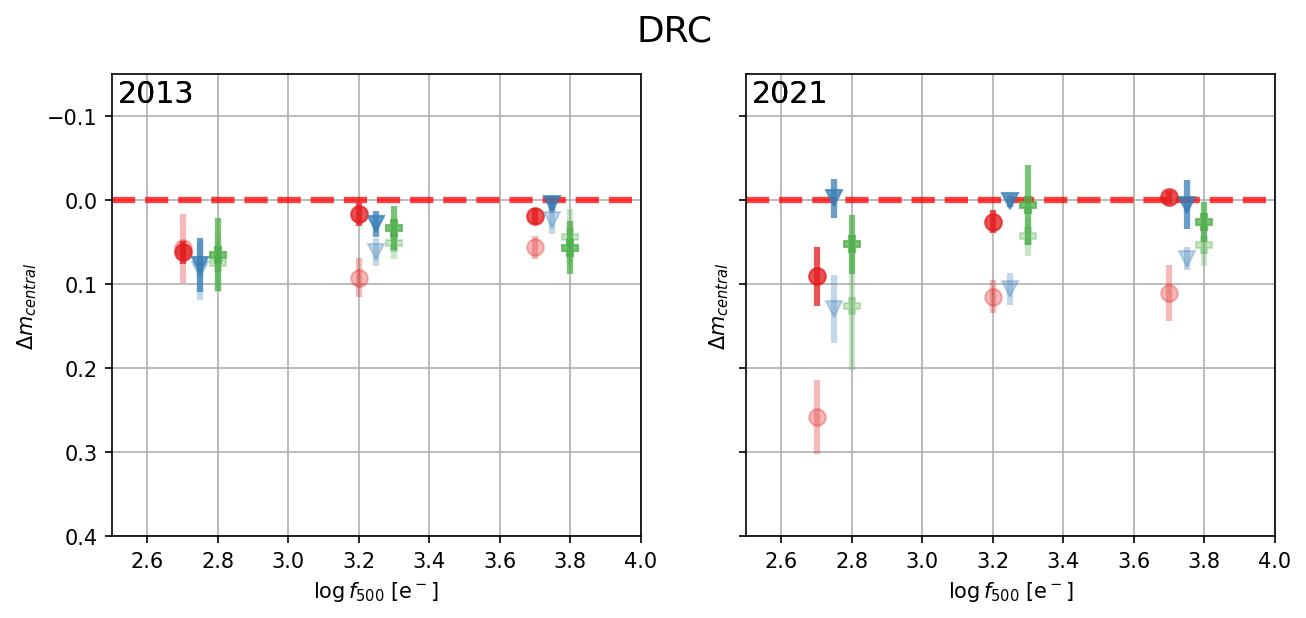}
    \caption{Same as Figure~\ref{fig:dmag_dy_fluence_kronfactor2} but with $\Delta m_{central}$ on the $y$ axis. As in Figure~\ref{fig:dmag_yshifts_background_kron0.5}, central brightness measurements are significantly more underestimated than global brightness measurements, especially for faint sources on low backgrounds far from the serial registers.}
    \label{fig:dmag_dy_fluence_kronfactor0.5}
\end{figure}

Figure~\ref{fig:dmag_dy_fluence_kronfactor0.5} shows the same analysis for $\Delta m_{central}$, where we see a clearer dependence on $f_{500}$ at fixed background and $\Delta y$ such that dimmer sources show stronger losses that can still be significant at low $\Delta y$. For the 2013 DRZ measurements at low $\Delta y$, these losses vary between 0.1 mags and 0.025 mags at the lowest and highest brightnesses, respectively, with no significant additional dependence on background. At high $\Delta y$, the losses roughly double, but again do not show a significant background dependence. The DRC measurements lower $\Delta m_{central}$, especially at large $\Delta y$. Correlations between $\Delta m$ and $\log{f_{500}}$ weaken below 3$\sigma$ significance, and brightness losses stay below 0.065 mags averaged across the detectors.

In the 2021 DRZ measurements at low $\Delta y$, only the data at backgrounds $<20e^-/{\rm pix}$ show a clear correlation with $\log{f_{500}}$. The data at higher backgrounds show average $\Delta m$ of $\sim0.05$ across all $f_{500}$ values. At higher $\Delta y$, losses exceed 0.1 mags at all backgrounds and $f_{500}$, surpassing 0.3 mags at $\log{f_{500}}<3\,e^-$. Using DRC measurements slightly improves brightness measurements at low $\Delta y$, although these still approach 0.1 mags for sources with $\log{f_{500}} < 3\,e^-$ and background $<20e^-/{\rm pix}$. At high $\Delta y$, losses exceed 0.1 mags at $\log{f_{500}} < 3\,e^-$, regardless of background. The brightest sources still have $\Delta m$ ranging from 0.05-0.1 this far from the serial registers. 

Based on this analysis, it is recommended that users ensure targets are placed at $\Delta y < 512$ if they require reliable spatially resolved brightness measurements. For objects with $\log{f_{500}}>3\,e^-$ and backgrounds $>30e^-/{\rm pix}$, different detector positions may be tolerable if systematic losses of 0.05-0.1 mags are acceptable.

\subsection*{A closer look at link between background, total recovered counts, and signal-to-noise}

Figure~\ref{fig:dmag_fluence} shows that below $\log{f_{500}}\sim2.5$ brightness recovery in later epoch data plummets dramatically, regardless of background level. However, it is useful to keep in mind that background and the number of recovered counts are linked, so objects that have $\log{f_{500}} < 2.5$ when at backgrounds $<20e^-/{\rm pix}$ may have $\log{f_{500}} > 2.5$ at higher backgrounds. In other words, data points move diagonally to the upper right in Figure~\ref{fig:dmag_fluence} as their backgrounds increase. 

\begin{figure}
    \centering
    \includegraphics[width=\linewidth]{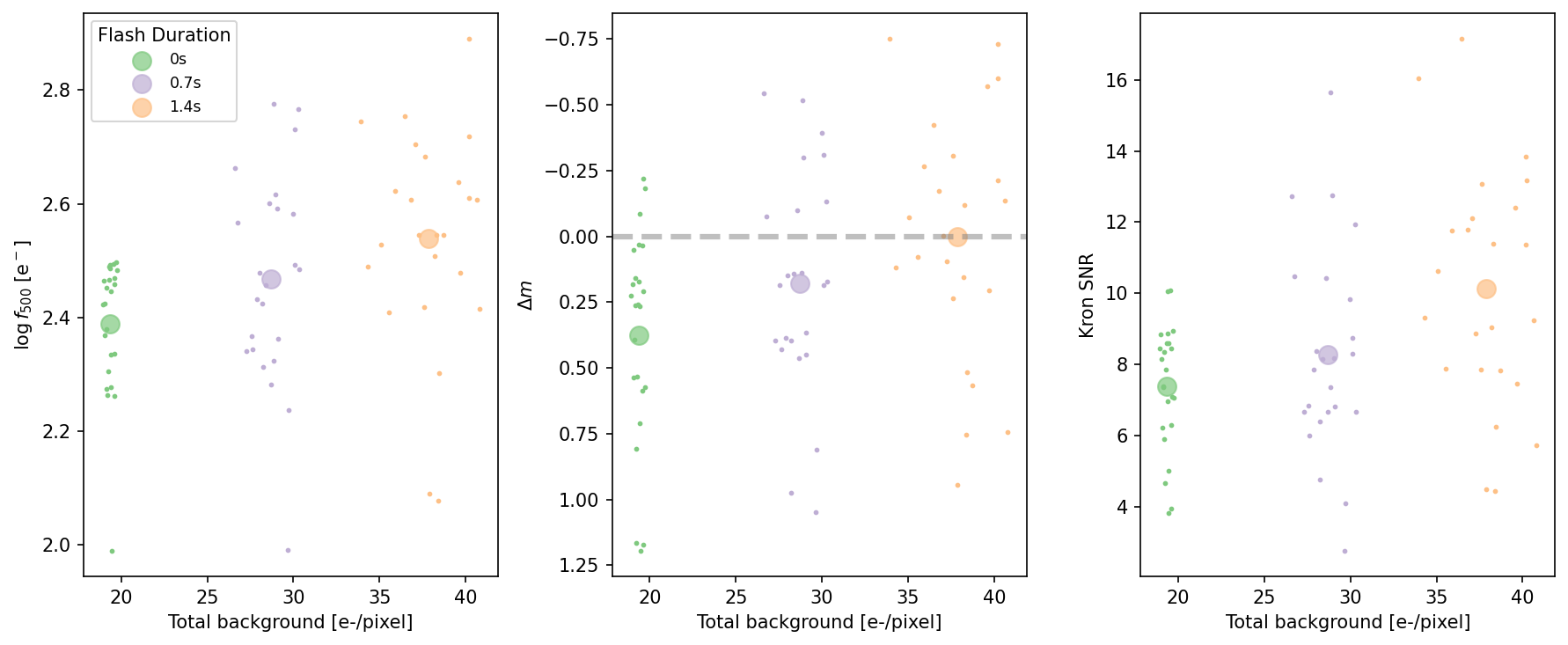}
    \caption{Total $f_{500}$, $\Delta m$, and $SNR$ as a function of background level for the subset of sources chosen to have 
    $\log{f_{500}}<2.5$ and $\Delta y > 1024$ in the unflashed 2021 data. Increasing the background level (in this case through post-flash) significantly increases the number of source counts and $SNR$, while decreasing $\Delta m$.}
    \label{fig:snr_dmag_weakest}
\end{figure}

 To demonstrate this point, we isolate sources in the non-post-flashed 2021 DRZ imaging data with $\log{f_{500}} < 2.5$ and $\Delta y>1024$ (the half of the detector furthest from the serial registers where CTI-related effects are most pronounced), which have backgrounds of $\sim19e^-/\mathrm{pixel}$ on average. We identify these exact same sources in the post-flashed 2021 data and plot $\log{f_{500}}$, $\Delta m$, and $SNR$ as a function of the local background in Figure~\ref{fig:snr_dmag_weakest}\footnote{This is distinct from our earlier analysis because we are tracking the exact same sources across all backgrounds. Our prior analysis selected anything with $SNR>3$ in a given image, but did not require $SNR>3$ in all available images.}. As the mean background increases from $\sim19$ to $\sim38e^-/\mathrm{pixel}$, the number of source counts per exposure increases by 40\% on average, the mean value of $\log{f_{500}}$ exceeds 2.5, and $\Delta m$ is approximately zero. Despite the added sky noise, the total SNR increases because more source counts are recovered, making the increase in sky noise less significant to the overall error budget. A similar result was shown for point-sources in \citep{Stark2024}, where SNR was shown to increase for simulated point sources up to a background level of $\sim 30e^-/\mathrm{pixel}$, after which declined again as the sky Poisson noise began to outweigh improved brightness recovery. Although we do not witness this turnover in our data set, we expect it would eventually occur for extended sources as well. 

The results of Figure~\ref{fig:snr_dmag_weakest} illustrate the benefit of keeping backgrounds above $\sim30e^-/\mathrm{pixel}$, especially when probing the very dimmest sources. In cases where this goal cannot be accomplished by adjusting exposure time, post-flash could be employed, as was done for this study. Keep in mind that the improvements from obtaining higher backgrounds may not be seen in sources that are brighter and/or closer to the serial registers, where CTI has less of an impact on their brightnesses. It is possible that increasing the background for these sources may actually lead to a decrease in SNR due to the added sky background noise but minimal change in their measured brightness.

\subsection*{Impact on two-dimensional structure}

Up to this point, we have only considered integrated brightness measured within different apertures. We have not addressed how the two-dimensional structure of galaxies may be changed in the presence of degraded CTE, although our findings that central magnitudes are more severely impacted than total magnitudes, even when DRC images are used, implies that degraded CTE is skewing these 2D distributions and the pixel-based CTE correction is not perfectly placing counts back where they came from. In WFC observations of point sources, degraded CTE clearly ``smears" counts in the direction opposite from the CCD serial registers, and a similar smearing effect is likely impacting extended sources as well.

\begin{figure}
    \centering
    \includegraphics[width=\linewidth]{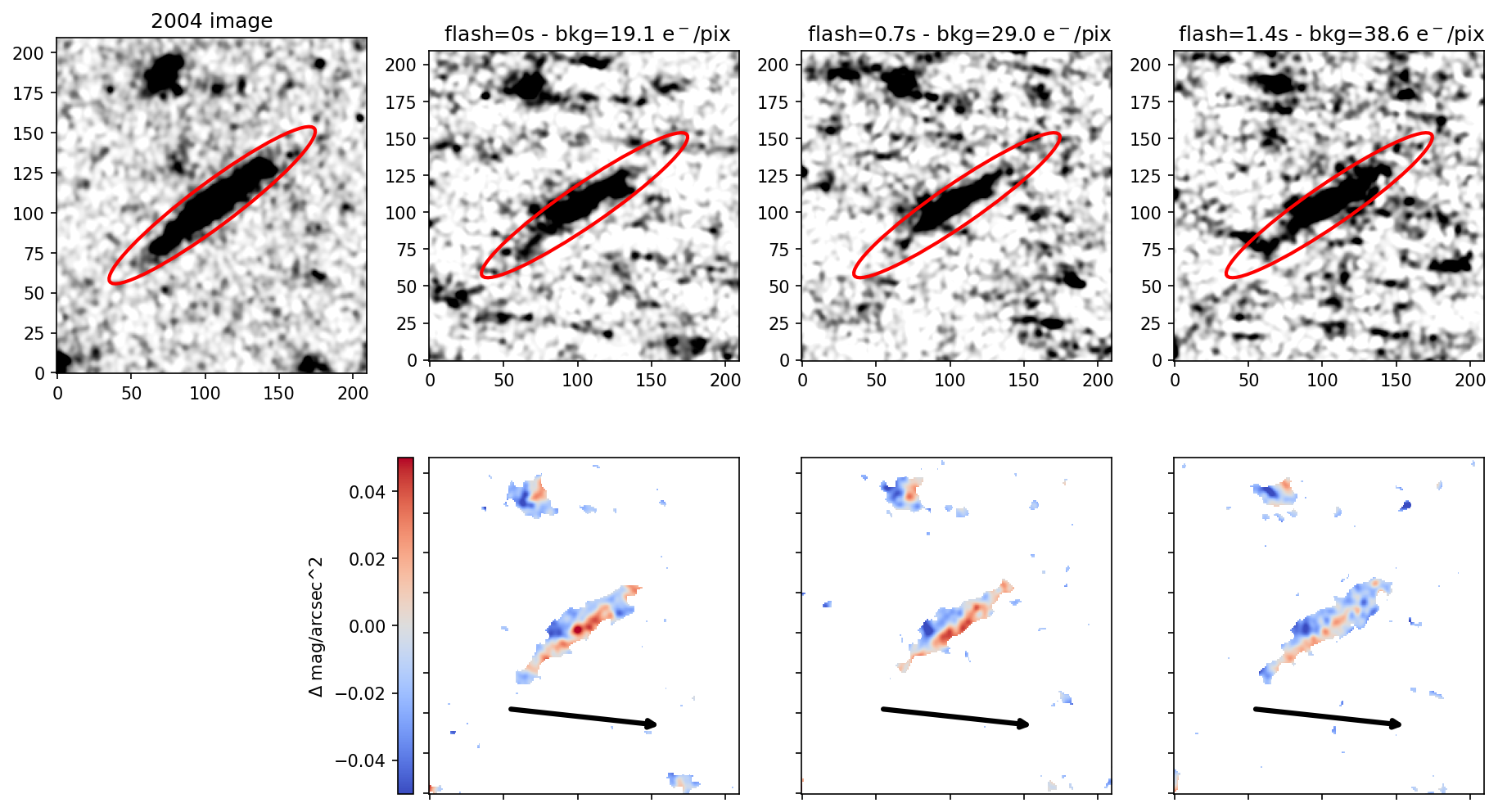}
    \includegraphics[width=\linewidth]{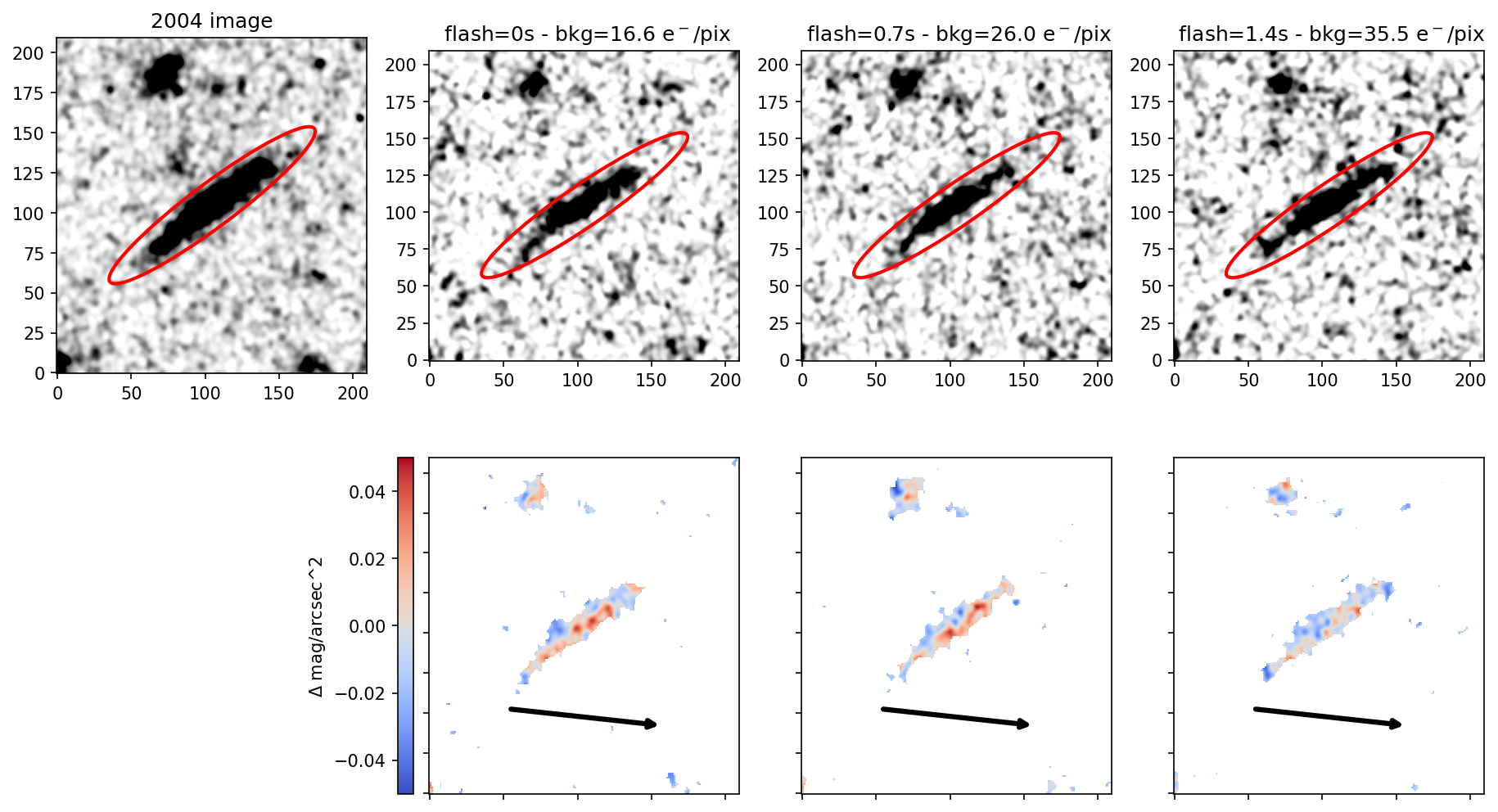}
    \caption{Cutouts and $\Delta \mu$ maps from DRZ (top) and DRC (images) for an example galaxy at $\Delta y > 1500$ pixels. In the top row, the left panel shows the ``truth" image from 2004, while the latter three images show the cutouts from the three visits from 2021 with different post-flash levels. The second row shows the difference in surface brightness, $\Delta \mu$, between these three visits and the 2004 data. The $\Delta \mu$ images are convolved with a Gaussian with FWHM of 4 pixels to reduce noise. Any pixels with absolute values $<2$ times the sky noise are masked. The arrow shows the approximate direction of the serial registers. This object shows a gradient in $\Delta \mu$ aligned with the readout vector.}
    \label{fig:dmag_ratio_highy}
\end{figure}

\begin{figure}
    \centering
    \includegraphics[width=\linewidth]{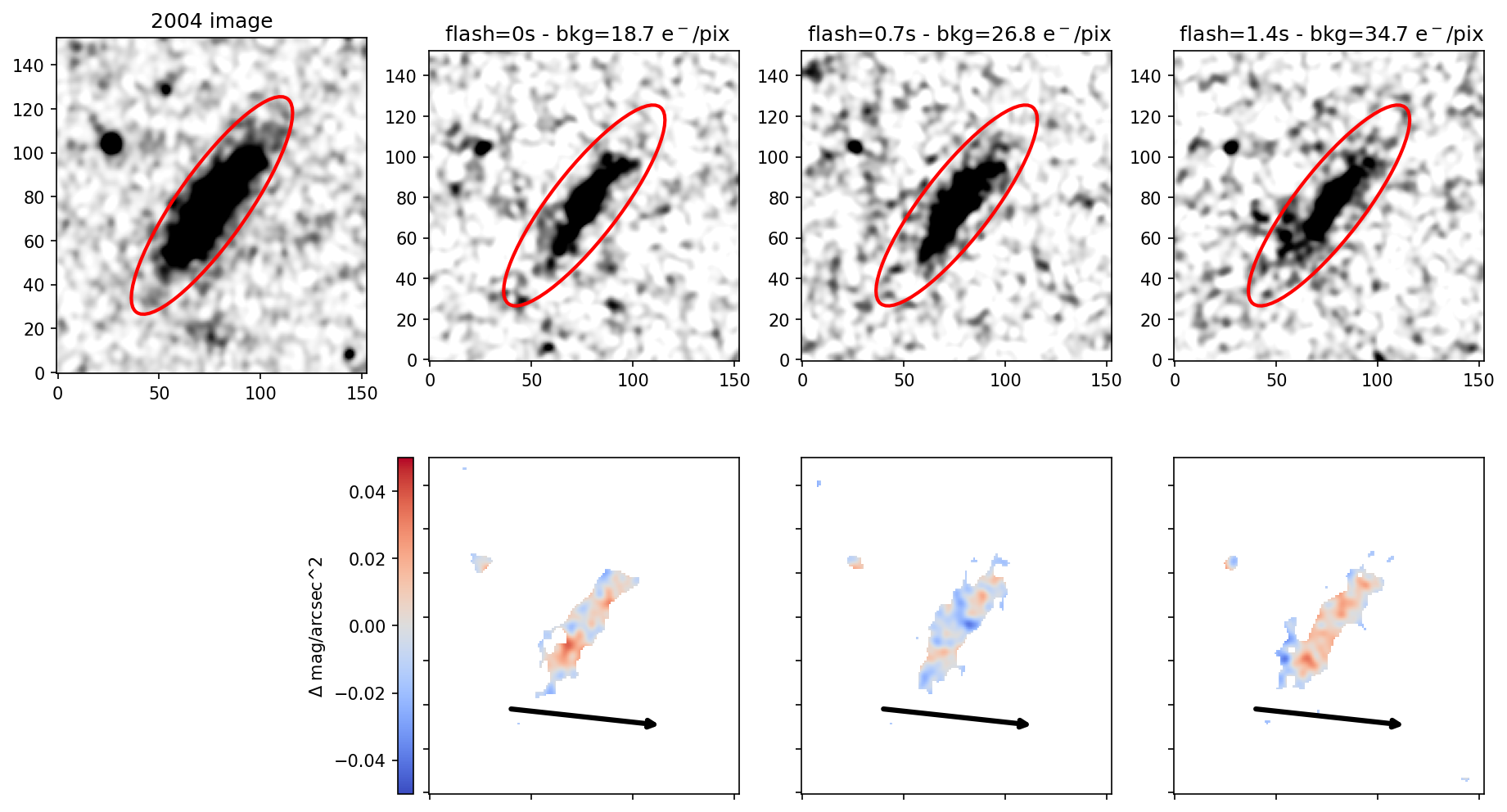}
    \includegraphics[width=\linewidth]{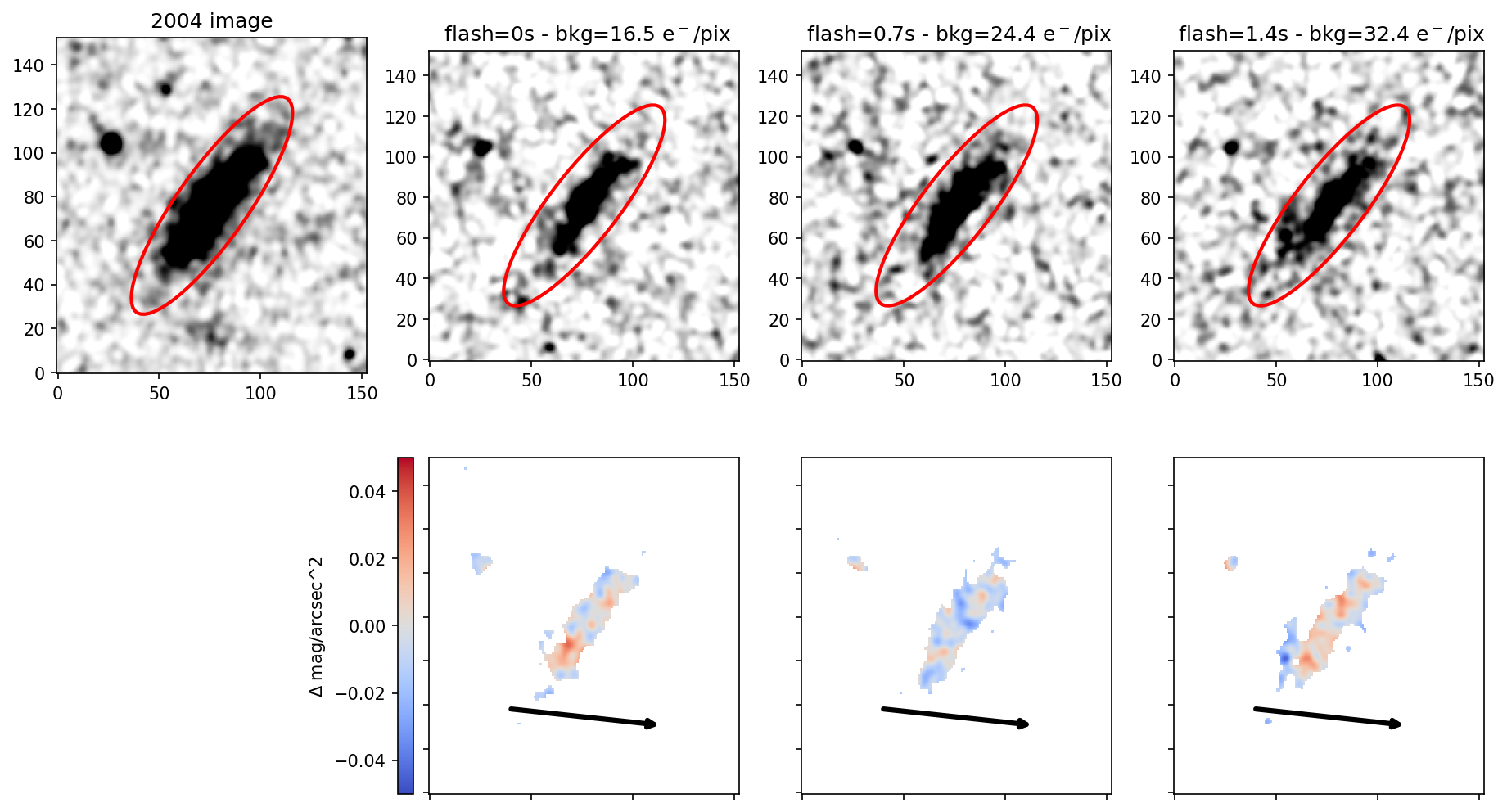}
    \caption{Same as Fig~\ref{fig:dmag_ratio_highy} but for a galaxy at $\Delta y < 500$ pixels. This object does not show a gradient in $\Delta \mu$ aligned with the readout vector.}
    \label{fig:dmag_ratio_lowy}
\end{figure}

To test this idea, we examine two-dimensional surface brightness ratios, $\Delta \mu$, which has the same definition as $\Delta m$ (Eq. 2) except that integrated magnitudes are replaced by surface brightness in units of magnitudes per arcsec$^2$. Figures~\ref{fig:dmag_ratio_highy} and \ref{fig:dmag_ratio_lowy} show examples of $\Delta \mu$ for two sources, one at $>1500$ pixels from the serial register, and one at $<500$ pixels from the serial register, i.e., one expected to be significantly impacted by degraded CTE, and one expected to be minimally impacted by degraded CTE. These figures show cutouts from the 2004 data and the 2021 data at different flash levels, followed by maps of $\Delta \mu$ for each of these three 2021 images and the 2004 ``truth" image. To improve clarity, the $\Delta \mu$ maps are convolved with a Gaussian kernel with FWHM of 4 pixels, and regions where the brightness is below two times the background noise level of either the 2004 image or the 2021 image are masked. The arrow overlaid on each $\Delta \mu$ map indicates the direction of the serial registers, i.e., the path charge takes during read-out\footnote{This vector can be shown because all four individual exposures share the same orientation.}. 


The source shown in Figure~\ref{fig:dmag_ratio_highy} shows a noticeable gradient in $\Delta \mu$ that is aligned with the read-out vector, especially in the DRZ images. Although not shown here, other objects at similarly large $\Delta y$ show this behavior as well. This gradient can be understood as the result of degraded CTE. As charge from the source is ``smeared" opposite the read-out direction, an excess ``behind" the galaxy and a deficit ``in front" of the galaxy forms. This gradient is substantially weaker in maps of $\Delta \mu$ created using DRC images. Additionally, the object close to the serial registers shown in Figure~\ref{fig:dmag_ratio_lowy}, which was chosen to have a similar shape, orientation, and brightness as the galaxy shown in Figure~\ref{fig:dmag_ratio_highy}, does not show a clear gradient aligned with the read-out direction in either the DRZ or DRC images at any post-flash level. 

Due to the lower SNR of the later epoch data used in our analysis, we can only examine two-dimensional maps of $\Delta \mu$ for the brighter/larger objects in our field. To analyze the average asymmetry of the whole population of sources in our field, we rely on stacking radial surface brightness profiles for all sources. The isophotes for these surface brightness profiles are spaced by 0.25 Kron radii so that galaxies of different absolute sizes can be stacked together. However, for each galaxy, we calculate the surface brightness profile in three different ways:
\begin{enumerate}
    \item Using the whole galaxy.
    \item Using only the half of the galaxy (divided along its major axis) furthest from the serial register.
    \item Using only the half of the galaxy (divided along its major axis) closest to the serial register.
\end{enumerate}

\begin{figure}
    \centering
    \includegraphics[width=0.5\linewidth]{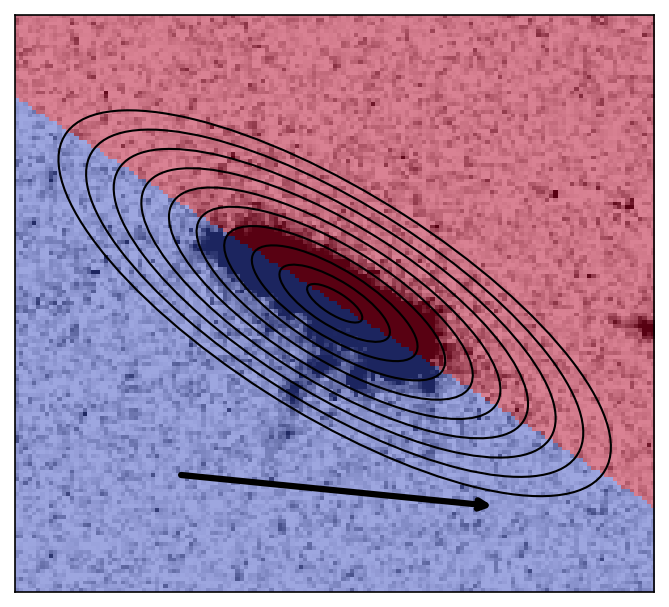}
    \caption{An example showing how radial profiles are calculated for our analysis of CTI-induced asymmetry. Surface brightnesses are measured using the full galaxy in fractions of Kron radii (black ellipses), but then the image is divided along the galaxy major axis into the side closer to the serial register (red) and further from the serial register (blue). Galaxies with PA values within $30^\circ$ of the read-out vector are not included in subsequent analyses because they do not have a clear "side" aligned with the read-out direction.}
    \label{fig:ro_masking_demo}
\end{figure}

The manner in which the galaxies are divided in half for this analysis is illustrated in Figure~\ref{fig:ro_masking_demo}. In some cases, a galaxy's position angle may be aligned closely with the read-out direction vector (which is approximately $6^{\circ}$ from the $+x$ axis), such that neither side, after dividing the galaxy along its major axis, is clearly ``facing" the serial registers. Therefore, we exclude any galaxies that have position angles within $30^{\circ}$ of the readout-direction vector. The general assumption behind this analysis is that if degraded CTE leads to systematic brightness gradients that align with the read-out direction, then we expect the surface brightness profiles measured on either side of our galaxies to be different, specifically dimmer and brighter on the sides closer to and further from the serial registers. Although galaxies often have natural, non-CTE related, asymmetries (e.g., \citealt{Conselice2000}), these should be randomly oriented and averaged-out when stacking multiple galaxies together, and so should not mimic any systematically aligned asymmetry driven by degraded CTE. 

\begin{figure}
    \centering
    \includegraphics[width=\linewidth]{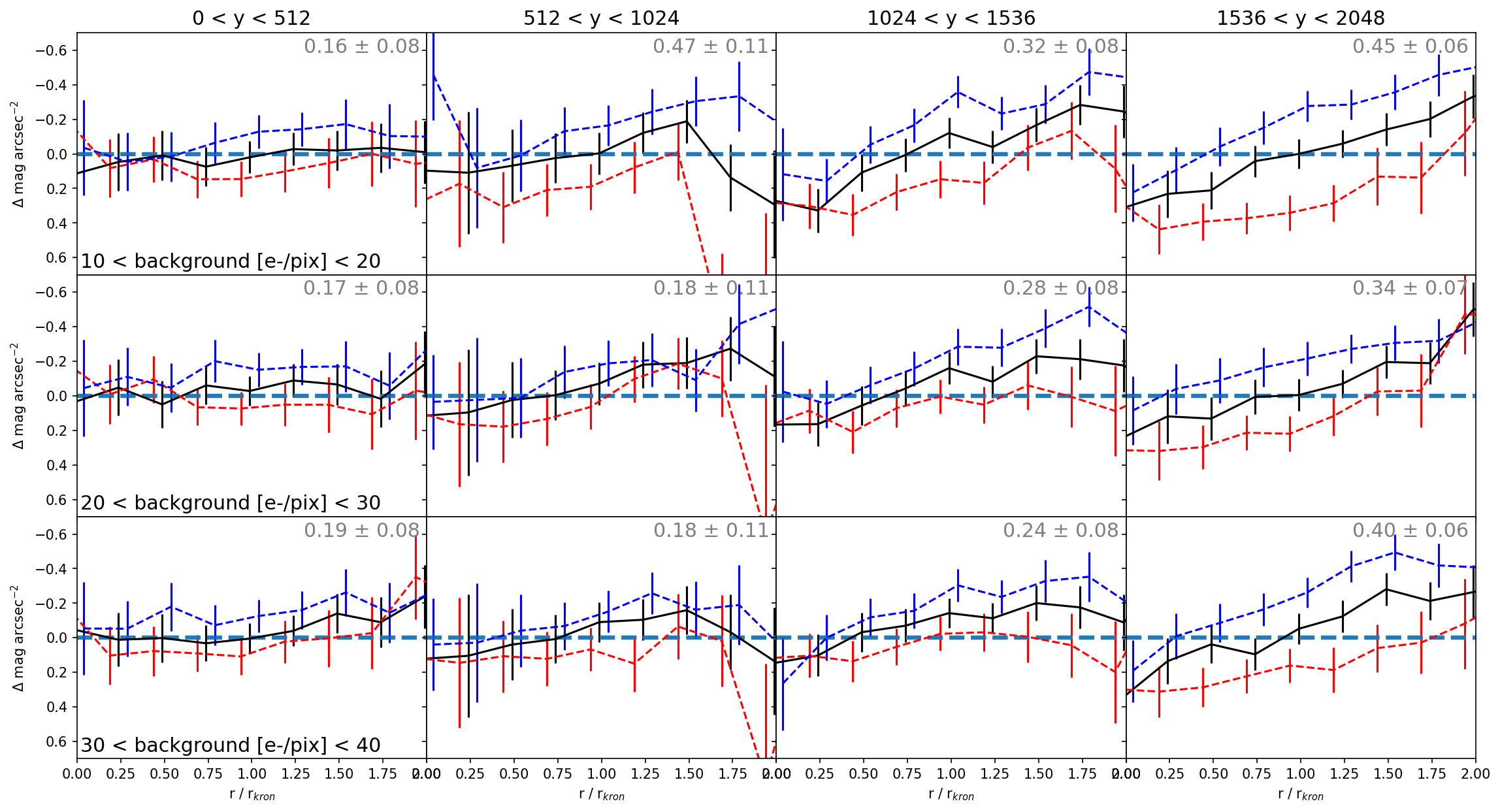}
    \caption{Mean profiles of $\Delta \mu$ measured in DRZ images and obtained using entire galaxies (black), just the sides closest to the readout direction (red), and just the sides furthest from the readout direction (blue). The difference in surface brightness measured on the readout and anti-readout directions   increases with increasing distance from the serial registers and decreasing background.}
    \label{fig:dmu_drz}
\end{figure}

Figure~\ref{fig:dmu_drz} shows a comparison of stacked radial profiles measured using the 2021 DRZ images as a function of background level and distance from the serial registers. For clarity, we do not plot surface brightness directly, but rather $\Delta \mu$ to emphasize differences between the 2021 and 2004 data. Black, blue, and red lines correspond to the radial profiles measured using the full galaxy, the side further from the serial register, and the side closer to the serial register. Each panel also prints in the upper right the average difference and its uncertainty between the radial profiles measured on either side of the galaxies. 

The average profiles using all the galaxy light (black lines) often show a brightness deficit at small radii, which is most pronounced at large $\Delta y$ and/or low background, and consistent with our earlier results studying $\Delta m_{central}$. Additionally, the surface brightness profiles measured on either sides of our galaxy sample are clearly distinct, with the further side typically showing a brightness excess and the closer side showing a brightness deficit, matching our expectations. The mean difference generally increases with increased $\Delta y$ and decreases with increased background. This mean difference is significant at $>3\sigma$ when $\Delta y>512$ pixels and background is $<20e^-/\mathrm{pixel}$, or when $\Delta y > 1024$ pixels at all backgrounds up to $\sim40e^-/\mathrm{pixel}$ probed by this study. 

\begin{figure}
    \centering
    \includegraphics[width=\linewidth]{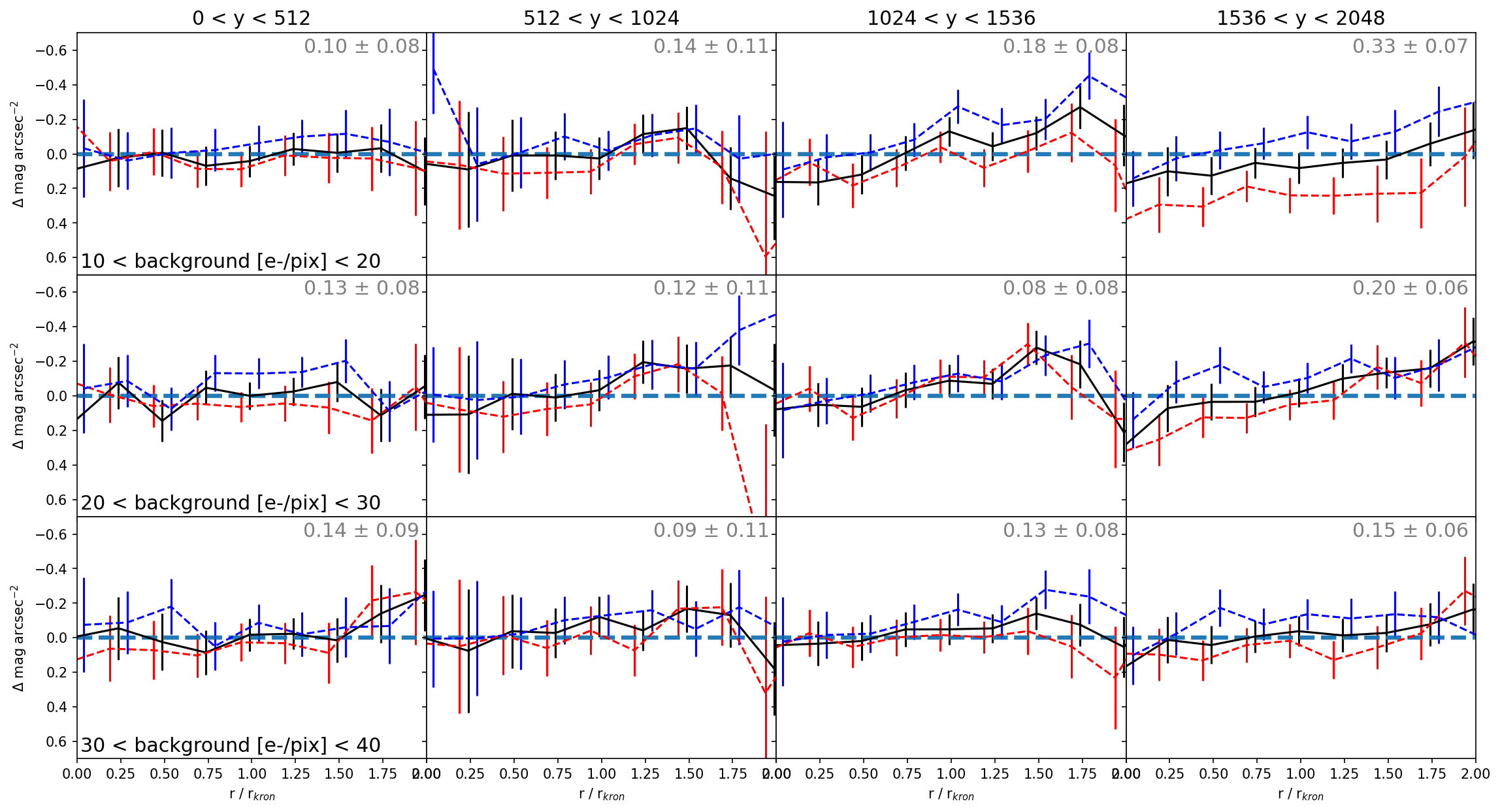}
    \caption{Same as Fig~\ref{fig:dmu_drz} except using DRC images. The difference between the measurements on the readout versus anti-readout sides of galaxies is significantly less over most parameter space.}
    \label{fig:dmu_drc}
\end{figure}

Figure~\ref{fig:dmu_drc} shows results from the exact same analysis except using the 2021 DRC images. Use of the DRCs notably improves the agreement between the surface brightness profiles measured on each side. The $\Delta \mu$ profiles using the full galaxies still show central deficits but are more flat indicating good agreement between the 2021 and 2004 data. There is still a $>3\sigma$ mean difference between the profiles on either side when $\Delta y>1536$ and background $<30e^-/{\rm pix}$, but otherwise the mean differences are much smaller, typically $<0.2\, {\rm mag\,arcsec^{-2}}$. The DRC image improves the asymmetry over most of the image, but users are advised to use caution when analyzing the 2D properties of sources at backgrounds $<20e^-/\mathrm{pixel}$ or at $\Delta y>1536$.

\section*{Discussion and Conclusions}
We have analyzed aperture-matched photometry and 2D images of sources in the galaxy cluster CL0024+16 across a baseline of seventeen years to assess the impact of degraded CTE on measurements of extended sources. By comparing the later epoch measurements to the minimally CTE-impacted 2004 data, we find the following:
\begin{itemize}
\item Total magnitudes using large apertures (with radius equal to twice the Kron radius) are recovered in later epoch data to within a systematic error of $\lesssim0.05$ mags (with no significant dependence on chip position) if backgrounds in individual exposures are $>20e-/\mathrm{pixel}$. At lower backgrounds, brightness measurements may carry larger discrepancies, especially for objects far from the CCD serial registers and with brightnesses of $\lesssim 1000\,e^-$ per exposure (Figures~\ref{fig:dmag_yshifts_background_kron2} and \ref{fig:dmag_dy_fluence_kronfactor2}).

\item Regardless of background level, total brightnesses become significantly underestimated once their brightness is $\lesssim 300\,e^-$ per exposure (Figure~\ref{fig:dmag_fluence}). 

\item Brightnesses measured using smaller apertures (with radius equal to half the Kron radius) that cover only the centers of galaxies show much stronger discrepancies at later epochs that steadily increase as sources get further from the serial registers. Systematic discrepancies can be kept to $\lesssim 0.05$ mags if individual exposure backgrounds are \mbox{$>20e^-/{\rm pix}$} and targets are less than $512$ pixels from the serial registers (Figures~\ref{fig:dmag_yshifts_background_kron0.5} and \ref{fig:dmag_dy_fluence_kronfactor0.5}) 

\item Degraded CTE induces artificial asymmetries into galaxies. These asymmetries are minimized when DRC images are used, exposure backgrounds are $>20e^-/\mathrm{pixel}$, and sources are less than $1536$ pixels from the serial registers (Figures~\ref{fig:dmu_drz} and \ref{fig:dmu_drc}). 
\end{itemize}

These results primarily reflect measurements using the 2021 DRC images. Brightness discrepancies tend to be more pronounced in DRZ images. Additionally, since CTE degradation increases with time, the discrepancies observed in the 2013 imaging data are smaller, though not non-zero. We urge readers to consult our figures carefully to determine how their own data may be affected. 

Based on our findings, we make the following recommendations to present-day WFC users:
\begin{itemize}
\item Use DRC images instead of DRZ images for scientific analysis. Although the pixel-based CTE correction is not perfect, brightnesses measured in DRCs show significantly better agreement across time than DRZs. 

\item Follow the current ACS/WFC recommendations and keep total background per WFC exposure to $>30e^-/\mathrm{pixel}$. If this cannot be achieved with the sky and dark current alone, the target of interest should be placed as close to the WFC CCD serial register as possible. If such placement is not possible, the use of LED post-flash should be considered.  Because the post-flash intensity is significantly variable across the WFC CCDs, it is generally recommended only for compact targets less than $\sim$10 pixels across \citep{Ogaz2014}. Post-flash recommendations may evolve, as the WFC CTE further degrades.

\item Ensure any science target has $>300\,e^-$ per WFC exposure.

\item If spatially resolved brightness measurements are needed, sources should be placed as close to the WFC CCD serial register as possible, ideally within $\sim 512$ pixels, especially for targets with $\lesssim 1000 e^-/{\rm pix}$. Larger distances may be acceptable for brighter objects if background recommendations are followed.
\end{itemize}

Our findings have implications for scientific studies. Generally speaking, underestimating source brightness will impact estimates of physical properties, e.g., SED shape and subsequently inferred parameters like stellar mass and star formation history. The underestimation of nuclear brightnesses may impact measurements such AGN brightness inside a host galaxy, as well as quantitative galaxy morphology metrics such as concentration and Sersic indices that rely on central brightness measurements. Additional morphology metrics may be impacted as well. We have already shown that degraded CTE can create artificial asymmetries, and its impact on other higher-order morphology indicators will be explored in future work.

There are important limitations to this study. We only investigated how parallel CTI (i.e., in the $y$ direction in detector space) impacts brightnesses. Additional inefficiencies exist when charge is transferred in the serial ($x$) direction, although these losses are expected to be significantly smaller than those in the parallel direction \citep{Ryon2024}. We focused on extended sources with size scales of 10s-100s of pixels in size, and have not explored observations of large targets that encompass a significant fraction of the WFC field of view (e.g., nearby galaxies, nebulae) although it is generally expected that such large objects will be ``self-shielded" from the worst effects of degraded CTE. It is also possible that brightness measurements impacted by degraded CTE depend on more than background, source brightness, and detector position. Other properties such as galaxy shape, orientation with respect to the read-out direction, and surface brightness may also play a role. With our limited sample size, we could not explore these properties in this study, but more elaborate multivariate analysis may be possible in the future using the extensive archival WFC data available. Nonetheless, our findings provide important guidance for how users should prepare their observations to obtain reliable scientific results. 

\section*{Acknowledgments}
We thank Ray Lucas, Jenna Ryon, and Gagandeep Anand for useful discussions and comments on this study. This work made use the \href{https://numpy.org/}{Numpy} \citep{numpy}, \href{https://www.astropy.org/index.html}{Astropy}  \citep{astropy:2013,astropy:2018,astropy:2022}, \href{https://matplotlib.org/stable/index.html}{Matplotlib} \citep{matplotlib}, \href{https://github.com/astropy/photutils}{photutils} \citep{photutils}, \href{https://scikit-image.org/}{scikit-image} \citep{scikit-image},
\href{https://scipy.org/}{scipy} \citep{scipy} and \href{https://drizzlepac.readthedocs.io/en/latest/}{Drizzlepac} \citep{drizzlepac1,drizzlepac2} Python packages.

\bibliography{miles_cmdtime}
\bibliographystyle{apj}

\end{document}